\documentclass[aps,prb,10pt,twocolumn,a4paper,superscriptaddress,longbibliography]{revtex4-2}

\usepackage[colorlinks = true,
            linkcolor = blue,
            urlcolor  = blue,
            citecolor = blue,
            anchorcolor = blue,
            unicode]{hyperref}

\usepackage{graphicx}
\usepackage{amsmath}
\usepackage{amssymb}
\usepackage{amsfonts}
\usepackage[dvipsnames]{xcolor}
\usepackage{soul}
\usepackage{float}

\DeclareMathOperator{\arccosh}{arccosh}

\begin{document}

\title{Superconducting orbital diode effect in SN bilayers}

\author{Yuriy~A.\ Dmitrievtsev}
\affiliation{L.~D.\ Landau Institute for Theoretical Physics RAS, 142432 Chernogolovka, Russia}
\affiliation{Moscow Institute of Physics and Technology, 141700 Dolgoprudny, Russia}

\author{Yakov~V.\ Fominov}
\affiliation{L.~D.\ Landau Institute for Theoretical Physics RAS, 142432 Chernogolovka, Russia}
\affiliation{Moscow Institute of Physics and Technology, 141700 Dolgoprudny, Russia}
\affiliation{Laboratory for Condensed Matter Physics, HSE University, 101000 Moscow, Russia}

\begin{abstract}
We study the superconducting diode effect (SDE) in a diffusive superconductor -- normal metal (SN) bilayer subjected to an in-plane magnetic field. The supercurrent flows along the layers, perpendicular to the field. The SDE, manifested as an asymmetry in the critical (depairing) currents and kinetic inductance for opposite current directions, arises from an orbital mechanism due to the inhomogeneous distribution of the Meissner currents caused by a spatially varying superfluid density. Recently, Levichev \textit{et al.} [Phys.\ Rev.~B \textbf{108}, 094517 (2023)] demonstrated the realization of this effect in such a structure, supporting numerical calculations for an ideal interface with an experiment.
In this work, we investigate the influence of a nonideal interface with finite resistance on the SDE. Employing an analytical approach, we focus on limiting cases corresponding to weak intralayer inhomogeneities. We find that the strength of the SDE depends nonmonotonically on the interface resistance when the bilayer thickness is small compared to the coherence length. Remarkably, a nonideal interface can enhance the SDE compared to the ideal case.
We also demonstrate that, in the vicinity of the phase transition, a regime of unidirectional superconductivity can be realized, in which superconductivity exists only within an interval of finite currents flowing in the same direction.
\end{abstract}

\date{27 September 2026}

\maketitle

\tableofcontents

\section{Introduction}
\label{sec:intro}

Nonreciprocal transport phenomena in superconducting systems have been known for a long time \cite{KulikYansonBook, BaroneBook, Moll2023, Levitov1985, Edelstein1996, Krasnov1997.PhysRevB.55.14486, Majer2003.PhysRevLett.90.056802, Villegas2003, Vodolazov2005.PhysRevB.72.064509, deSouzaSilva2006, Morelle2006, Aladyshkin2010.10.1063/1.3474622, Silaev2014, Yokoyama2014.PhysRevB.89.195407, Wakatsuki2017, Chen2018.PhysRevB.98.075430}.
Recently, this area of research, now commonly known as the superconducting diode effect (SDE), has attracted renewed and significant attention from both fundamental and applied perspective (see Refs.\ \cite{Nadeem2023,Nagaosa2024,Shaffer2025arXiv} for recent reviews covering theoretical and experimental results in a variety of physical platforms and due to various physical mechanisms).
A clear manifestation of the SDE is an asymmetry in current transport: the critical supercurrents are different for two opposite directions along a given structure.

An essential ingredient for such asymmetry is a symmetry breaking. Typically, this requires a specific direction for current transport, combined with the breaking of inversion and time-reversal symmetries.

As recently pointed out by Levichev \textit{et al.} \cite{Levichev2023.PhysRevB.108.094517}, a promising platform for observing this effect is a superconductor (S) / normal metal (N) bilayer subjected to a parallel magnetic field. Due to the proximity effect, a superfluid density gradient $\nabla n$ emerges in the direction perpendicular to the bilayer's plane. This gradient, in conjunction with the applied magnetic field $\mathbf{B}$, defines a specific direction $[\nabla n \times \mathbf{B}]$ for the applied current $\mathbf{I}$, as illustrated in Fig.~\ref{fig:bilayr}. Furthermore, the magnetic field breaks the time-reversal symmetry. The presence of these conditions makes the SDE possible in such a structure. 
In the considered structures, which are not atomically thin, the magnetic field primarily influences the orbital motion of electrons, allowing the Zeeman energy to be neglected. Thus, the effect is of purely orbital nature (due to a nontrivial distribution of Meissner currents). In addition to asymmetry in the critical (depairing) currents, the SDE manifests itself in an asymmetry of the kinetic inductance with respect to current direction reversal, $L_k(I) \neq L_k(-I)$ \cite{Baumgartner2022, Levichev2023.PhysRevB.108.094517,Nadeem2023}.

Levichev \textit{et al.} \cite{Levichev2023.PhysRevB.108.094517} have studied the SDE in a diffusive SN bilayer both theoretically and experimentally (in MoN/Cu bilayers). Theoretically, numerical calculations were performed for the case of layer thicknesses $d$ on the order of a few coherence lengths $\xi$ in the dirty limit ($l\ll\xi$, where $l$ is the mean free path), allowing the application of the Usadel equations \cite{Usadel1970,Belzig1999}. The case of a transparent interface was considered. When the layer thicknesses are on the order of $\xi$, this leads to a nontrivial spatial distribution of the superfluid density (density of the superconducting electrons) $n(x)$ across the bilayer, which generally renders the problem analytically intractable.

Analytical solutions are only feasible in cases of weak intralayer inhomogeneities (i.e., weak inhomogeneity within each layer). For a transparent interface, such weak inhomogeneity is achieved in a thin bilayer (i.e., at $d \ll \xi$). At the same time, weak inhomogeneity can also be realized at low interface transparency. This is because increased interface resistance weakens both the direct and inverse proximity effect, making the layers less sensitive to each other. Consequently, even for relatively thick bilayers, a sufficiently low interface transparency can yield the weak inhomogeneity, necessary for analytical treatment.

For the orbital mechanism of the SDE, the gradient $\nabla n$ of the superfluid density is necessary but not sufficient.
Importantly, the SDE requires not just a gradient, but rather a nontrivial functional dependence of the $n(x)$ \emph{profile} on the current $I$ and the magnetic field $B$, one that does not reduce to an overall amplitude variation.
Decreasing the interface transparency (increasing its resistance) creates additional inhomogeneity, as it leads to a discontinuity of the Green function. This, in turn, modifies the $n(x)$ profile and can enhance $\nabla n$. Thus, this additional inhomogeneity can enhance the SDE. On the other hand, in the limit of very low transparency, the SDE should disappear since the proximity-induced superconductivity in the N layer is strongly suppressed, leading to effective isolation of the S layer. The interplay between these two tendencies opens up the possibility of a nonmonotonic behavior of the SDE as a function of the interface transparency.
In this paper, we consider the influence of finite interface transparency on the SDE and demonstrate that a nonmonotonic behavior can indeed be realized.

The paper is organized as follows. 
In Sec.\ \ref{sec:general}, we formulate the model and our theoretical approach to studying the regime of weak intralayer inhomogeneities. 
In Sec.\ \ref{sec:Weak_case}, we consider the limit of a weakly nonideal interface in the case of a thin bilayer.
In Sec.\ \ref{sec:Strong_case}, we consider the limit of a strongly resistive interface.
In Sec.\ \ref{sec:AGlimits}, we provide a further analysis of the effective Abrikosov-Gor'kov-type equations obtained in Secs.\ \ref{sec:Weak_case} and \ref{sec:Strong_case}; specifically, we consider the limits of zero temperature and the vicinity of the phase transition.
In Sec.\ \ref{sec:thick}, we discuss extending the results of Sec.\ \ref{sec:Weak_case} to moderately thick bilayers.
Finally, we present our conclusions in Sec.\ \ref{sec:conclusion}. Some details related to the kinetic inductance and general details of the Abrikosov-Gor'kov theory are provided in Appendices.
Throughout the paper, we use the units with $\hbar = k_B = c = 1$.

\section{Method}
\label{sec:general}

\subsection{Main equations}
\label{eq:Main_eqs}

The superconducting proximity effect in SN bilayers has been studied for a long time \cite{Cooper1961.PhysRevLett.6.689,deGennes1964.RevModPhys.36.225,McMillan1968,Golubov1989,Golubov1994,Belzig1999,Fominov2001.PhysRevB.63.094518}. In the diffusive (dirty) limit, the system can be described with the help of the Usadel equation \cite{Usadel1970,LarkinOvchinnikov1986NoneqScReview,Belzig1999}. Our consideration will follow the approach of Ref.\ \cite{Fominov2001.PhysRevB.63.094518}, taking into account the in-plane magnetic field $B$ and the current $I$ along the bilayer.

\begin{figure}[t]
 \includegraphics[width=\columnwidth]{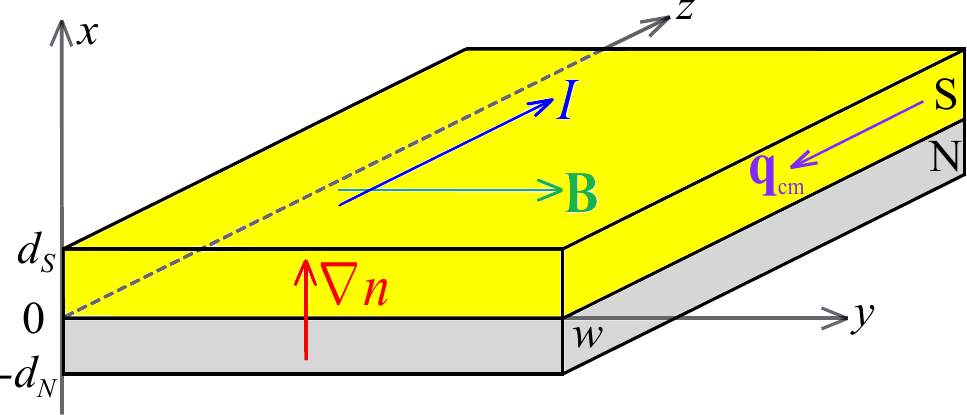}
 \caption{SN bilayer in the form of a strip in the in-plane magnetic field $\mathbf{B}$. Due to the proximity effect, a nonuniform distribution of the superfluid density, $n(x)$, arises across the bilayer thickness. The magnetic field breaks the time-reversal symmetry. We consider currents $I$ flowing along and opposite to the symmetry-breaking direction defined by $[\nabla n \times \mathbf{B}]$ (the $z$ axis). The SDE manifests as an asymmetry of the critical (depairing) currents in the positive and negative direction, $I_{c+} \neq I_{c-}$, and a nonreciprocal kinetic inductance, $L_k(I) \neq L_k(-I)$  \cite{Vodolazov2005.PhysRevB.72.064509}. The combined influence of the magnetic field and the current leads to an inhomogeneous momentum of the superconducting condensate, $\mathbf{q}_s(x)$, along the $z$ axis. The overall effect can be characterized by a center-of-mass momentum $\mathbf{q}_\mathrm{cm}$.} 
 \label{fig:bilayr}
\end{figure}

The system under consideration is a bilayered strip of width $w$ consisting of a superconducting layer ($0<x<d_S$) and a normal-metal layer ($-d_N<x<0$) with proximity-induced superconductivity, as illustrated in Fig.~\ref{fig:bilayr}. The layers are in the dirty limit, which implies that the mean free path is much smaller than superconducting and geometric scales. At the same time, the thicknesses of the layers are assumed to be much smaller than the London penetration depths
\begin{equation}
        \lambda_{S(N)} = \sqrt{m/4\pi n_{S(N)} e^2},
\end{equation}
where $n_{S(N)}$ is the superfluid densities in the layers. This implies full penetration of the magnetic field inside the bilayer.

We employ the angular $\theta,\varphi$ parametrization \cite{Zaikin1981,Belzig1999} of the Green function within the Usadel equation. The Usadel equations then take the form
\begin{gather}
    \frac{D}{2} \nabla ^2 \theta - \left(\omega_n + \frac{D}{2} \mathbf{q}_s^2 \cos \theta \right) \sin\theta + \Delta \cos \theta = 0, 
    \label{eq:usadel_first_appear}\\
    \nabla( \mathbf{q}_s \sin^2 \theta ) =0.
    \label{eq:trivial_for_phase}
\end{gather}
Here, the spectral angle $\theta$ characterizes the strength of superconductivity, $D$ is the diffusion constant, $\omega_n = \pi T (2n + 1)$ are the Matsubara frequencies at temperature $T$, $\Delta$ is the absolute value of the order parameter [referred to as ``the order parameter'' below for brevity, and which should be determined from self-consistency equation, see Eq.\ \eqref{eq:self_consist_initial} below], $\mathbf{q}_s = \nabla \varphi + 2 e \mathbf{A}$ is the gauge-invariant momentum of superconducting electrons, $\varphi$ is the phase of the order parameter, and $\mathbf{A}$ is the vector potential of the magnetic field (we choose $e>0$, so the electron charge is $-e$).
The orbital effect of the magnetic field thus enters the equations via $\mathbf{q}_s$. At the same time, we neglect the Zeeman term which would become essential only in the limit of extremely thin (of atomic thickness) layers. 
Our consideration implies that the thicknesses of the layers are much larger than the atomic scale.

The geometry of the system is such that all variations in the spectral angle occur solely in the direction perpendicular to the layers, i.e., $\theta = \theta(\omega_n, x)$. The momentum and the order parameter exhibit similar dependences, varying only along the $x$-axis: $\mathbf{q}_s =\mathbf{q}_s(x)$, $\Delta = \Delta(x)$. 
At the same time, Eq.\ \eqref{eq:trivial_for_phase} is automatically satisfied since  $\mathbf{q_s}(x)$ is directed along the $z$ axis. The Usadel equations are then reduced to Eq.\ \eqref{eq:usadel_first_appear} in each layer:
\begin{gather}
        \frac{D_S}{2} \partial_x ^2 \theta_S - \left(\omega_n + \frac{D_S}{2} \mathbf{q}_s^2 \cos \theta_S \right) \sin\theta_S + \Delta \cos \theta_S = 0, \label{eq:Usadel_first_1}
        \\
        \frac{D_N}{2} \partial_x ^2 \theta_N - \left(\omega_n + \frac{D_N}{2} \mathbf{q}_s^2 \cos \theta_N \right) \sin\theta_N  = 0.
        \label{eq:Usadel_first_2}
\end{gather}
While $\Delta(x) = 0$ in the N layer, the proximity-induced superconducting correlations are described by nonzero $\theta_N$.

The sample is subjected to the external magnetic field $\mathbf{B}$ applied along the $y$ axis. Given that the bilayer thickness $d$ is much smaller than the London penetration depth (i.e., $d \ll \lambda$), the magnetic field is constant inside the bilayer: $\mathbf{B} = \mathrm{const}$. From equation $\nabla \times \mathbf{q}_s = 2 e \mathbf{B}$, we then obtain 
\begin{equation}
    q_s(x) = q_0 - 2 e B x. 
    \label{eq:super_velosity_definition}
\end{equation}
Here, $q_0$ is the momentum at the interface, which is determined by the condition that the total current is equal to $I$ [see Eq.\ \eqref{eq:current_general} below].

The Usadel Eqs.\ \eqref{eq:Usadel_first_1} and \eqref{eq:Usadel_first_2} must be supplemented with the Kupriyanov-Lukichev boundary conditions \cite{Kupriyanov1988}:
\begin{gather}
        \sigma_S \partial_x \theta_S (0) =  \sigma_N \partial_x \theta_N (0) = r_B^{-1} \sin \bigl( \theta_S(0) - \theta_N(0) \bigr),\label{eq:boundary_def1} \\
        \partial_x \theta_S(d_S) =0, \quad
        \partial_x \theta_N(-d_N)=0. 
        \label{eq:boundary_def}
\end{gather}
Here, $r_B$ is the interface resistance of unit area, $\sigma_{S(N)}$ is the normal-state conductivity of the corresponding layer.

The order parameter $\Delta$ entering the Usadel equation should itself be found from the self-consistency equation 
\begin{equation}
    \Delta(x)  \ln \frac{T}{T_{cS}} = 2 \pi T \sum_{\omega_n>0} \left(\sin \theta_S(\omega_n, x)-\frac{\Delta(x)}{\omega_n}\right), \label{eq:self_consist_initial}
\end{equation}
where $T_{cS}$ is the critical temperature of a single S layer in the absence of a magnetic field and current. Equations \eqref{eq:Usadel_first_1}, \eqref{eq:Usadel_first_2}, \eqref{eq:boundary_def1}-\eqref{eq:self_consist_initial} determine functions $\theta_{S(N)}(\omega_n, x)$ and $\Delta(x)$ at a given $q_0$ in Eq.\ \eqref{eq:super_velosity_definition}. 

To obtain the complete solution, we need to determine $q_0$ from the condition that the total current is equal to $I$. In order to write the expression for the current, we introduce the superfluid densities in the two layers:
\begin{equation}
    n_{S(N)}(x) = \frac{2 \pi m \sigma_{S(N)} T }{e^2} \sum_{\omega_n>0} \sin^2 \theta_{S(N)}(\omega_n,x).
    \label{eq:plotnost}
\end{equation}
The total current can then be written as
\begin{equation}
    I = -\frac{ew}{2m}\int_{-d_N}^{d_S} n(x) q_s(x) dx.
    \label{eq:current_initial}
\end{equation}
Thus, the task is reduced to solving Eqs.\ \eqref{eq:Usadel_first_1}-\eqref{eq:current_initial}.

It is instructive to rewrite Eq.\ \eqref{eq:current_initial} in terms of integral quantities: full thickness $d = d_S + d_N$, average superfluid density $\bar n = \int n(x)dx/d$, and the center-of-mass momentum $q_{\mathrm{cm}}=q_s(x_\mathrm{cm}) = q_0 - 2 e Bx_\mathrm{cm}$, where the center-of-mass coordinate is $x_\mathrm{cm} = \int x n(x)dx/\bar n d$. The current can then be written as
\begin{equation}
    I = -\frac{e w d}{2m} \bar n(q_{\mathrm{cm}},B)q_{\mathrm{cm}}.
    \label{eq:current_general}
\end{equation}
Equation \eqref{eq:current_general} defines the relationship between the total current $I$ and the momentum of the center of mass $q_\mathrm{cm}$ \footnote{To avoid confusion, we note that in Ref.\ \cite{Levichev2023.PhysRevB.108.094517}, the quantity $q_{0}^{\text{\cite{Levichev2023.PhysRevB.108.094517}}}$, chosen as an independent variable characterizing the distribution of the condensate momenta inside the SN bilayer, is the thickness-averaged momentum: $q_0^{\text{\cite{Levichev2023.PhysRevB.108.094517}}} =\int q_s(x)dx/d$. At $B\neq 0$ and $I=0$, the current distribution is such that $q_0^{\text{\cite{Levichev2023.PhysRevB.108.094517}}}\neq 0$, therefore this state was referred to as the ``finite-momentum'' superconductivity. In contrast, we find it more physically transparent to use the center-of-mass momentum $q_\mathrm{cm}$ as the independent variable. By definition, $q_\mathrm{cm}=0$ at $I=0$. This, however, is only a technical difference in the description; physically, we are considering the same state termed finite-momentum superconductivity in Ref.\ \cite{Levichev2023.PhysRevB.108.094517}.
}.
To find the critical currents, it is convenient to consider $q_{\mathrm{cm}}$ and $B$ as independent variables. The critical currents are then obtained by maximizing and minimizing $I(q_{\mathrm{cm}},B)$ with respect to $q_\mathrm{cm}$ at the fixed magnetic field: $I_{c+}(B) = \max\limits_{q_{\mathrm{cm}}} I(q_{\mathrm{cm}},B)$,
$I_{c-}(B) = - \min\limits_{q_{\mathrm{cm}}}  I(q_{\mathrm{cm}},B)$.

The SDE, manifested as $I_{c+}(B)\neq I_{c-}(B)$, arises because the average superfluid density is not an even function of the momentum:
\begin{equation}
    \bar n(q_{\mathrm{cm}},B) \neq \bar n(-q_{\mathrm{cm}},B).
    \label{eq:diode_criteria}
\end{equation}
At the same time, the full time-reversal symmetry implies the following relations: $\bar n(q_{\mathrm{cm}},B) = \bar n(-q_{\mathrm{cm}},-B)$, $I(q_{\mathrm{cm}},B) =-I(-q_{\mathrm{cm}},-B)$, and, consequently, $I_{c+}(B) = I_{c-}(-B)$.

To quantify the effect, we introduce two different characteristics for the ``strength'' of the SDE. The first one is the diode efficiency:
\begin{equation}
    \eta(B) = \frac{|I_{c+}-I_{c-}|}{I_{c+}+I_{c-}},
    \label{eq:kontrast}
\end{equation}
and the second one is the normalized nonreciprocity:
\begin{equation}
    \widetilde{\eta}(B) = \frac{|I_{c+}-I_{c-}|}{2 I_{c0}},
    \label{eq:absolut_diod}
\end{equation}
where $I_{c0} \equiv I_{c+}(B=0) = I_{c-}(B=0)$ is the critical current in the absence of a magnetic field. As we will see below, $I_{c\pm}$ acquire linear corrections at the small magnetic field. Therefore, $\eta$ and $\widetilde \eta$ behave similarly at the small magnetic fields, $\eta \approx \widetilde \eta \propto B$. 

Another important physical quantity that can be measured experimentally is the kinetic inductance per
unit length of the strip \cite{Levichev2023.PhysRevB.108.094517}:
\begin{equation}
    L_k(I,B) = -\frac{1}{2e} \left(\frac{dI}{dq_0}\right)^{-1}.
    \label{eq:kinetic_appearance}
\end{equation}
The derivation and discussion of this formula are given in  Appendix~\ref{Appendix:Kinetic inductance}. The SDE manifests itself as a nonreciprocal current-dependent kinetic inductance, $L_k(I) \neq L_k(-I)$ \cite{Baumgartner2022, Levichev2023.PhysRevB.108.094517,Nadeem2023}.

\subsection{Limit of weak intralayer inhomogeneity}
\label{sec:weakdinhom}

The problem formulated in Sec.\ \ref{eq:Main_eqs} admits an analytical treatment only in the case of weak inhomogeneity of the solution \emph{within each layer}. 
We will expand the solution according to this small intralayer inhomogeneity, following the approach of Ref.\ \cite{Fominov2001.PhysRevB.63.094518} 
(at the same time, the solution \emph{in the whole bilayer} may be strongly inhomogeneous due to an arbitrary jump at the interface).
The exact applicability condition for this approximation depends on the additional parameter, the interface resistance. For the two limiting cases of small and large resistance, we discuss it in Secs.~\ref{sec:Weak_conditions} [see Eqs.\ \eqref{eq:appl_weak_2} and \eqref{eq:condition_Thouless}] and \ref{sec:Strong_conditions} [see Eqs.\ \eqref{eq:condition2} and \eqref{eq:tauEth}], respectively. 
Qualitatively, each layer must be thin enough so that the diffusion time across it is much smaller than the larger of the escape time from the layer and the coherence time $\Delta^{-1}$.

Under the above conditions, the solution of Usadel Eqs.\ \eqref{eq:Usadel_first_1} and \eqref{eq:Usadel_first_2} in each layer can be approximated by a quadratic function of $x$.
From the boundary conditions \eqref{eq:boundary_def}, we obtain the solutions in the following form: 
    \begin{gather}
        \theta_{S(N)}(x) = \theta_{S(N)0} \pm \delta \theta_{S(N),d}\left[
            1 - \bigg(1\mp \frac{x}{d_{S(N)}}\bigg)^2
            \right],
            \label{eq:ththdth}\\
            n_{S(N)}(x) = n_{S(N)0} \pm \delta n_{S(N),d}\left[
            1 - \bigg(1\mp \frac{x}{d_{S(N)}}\bigg)^2
            \right].
            \label{eq:nndn}
    \end{gather}
Here, $\theta_{S(N)0}= \theta(x=\pm0)$ and $n_{S(N)0} = n(x=\pm0)$ are the spectral angles $\theta$ and the superfluid densities at the interface in the corresponding layer, while $\delta \theta_{S(N),d}$ and $\delta n_{S(N),d}$ are values characterizing the corrections due to inhomogeneities associated with the finite thickness. We call them $d$ corrections and $d$ inhomogeneities (hence, the $d$ index);
they originate from intralayer inhomogeneity \emph{within} each layer.

In principle, inhomogeneity of the order parameter should also be taken into account: $\Delta(x) = \Delta + \Delta_1(x)$. However, we show further that the leading order $\Delta(x) = \Delta = \mathrm{const}_x$ is sufficient in the weakly inhomogeneous regime, while spatially-dependent higher-order corrections $\Delta_1(x)$ yield only minor contributions.

The densities $n_{S(N)0}$ and $d$ corrections $\delta n_{S(N),d}$ can be obtained by expanding Eq. \eqref{eq:plotnost} using the approximation $\theta_{S(N)} = \theta_{S(N)0}\pm \delta \theta_{S(N),d}$. 

To find the unknown coefficients $\delta \theta_{S(N),d}$, we integrate the Usadel equations over the thickness of the bilayer taking into account the boundary conditions
\eqref{eq:boundary_def1} and 
\eqref{eq:boundary_def}.
This yields
\begin{equation} \label{eq:soultions_neodnor}
        \delta \theta_{S(N),d} = ( r_{S(N)}/2 r_B ) \sin \theta_j,
\end{equation}
where $r_{S(N)} = d_{S(N)}/\sigma_{S(N)}$ represents the transverse resistance of the corresponding layer of unit area, and $\theta_j = \theta_{S0} - \theta_{N0}$ is the jump of the spectral angle at the interface.

The interface values $\theta_{S(N)0}$ should be found from the following equations:
\begin{align}
        -\frac{\sin \theta_j}{\tau_S\sin\theta_{S0} \cos\theta_{S0}} &= \frac{ \omega_n }{\cos\theta_{S0}}  + E_S - \frac{\Delta}{\sin\theta_{S0}} ,
        \label{eq:main_eq_neproz1}\\
        \frac{\sin \theta_j}{\tau_N \sin\theta_{N0} \cos\theta_{N0}} &= \frac{ \omega_n}{\cos\theta_{N0}} + E_N.
    \label{eq:main_eq_neproz}
\end{align}
The $E_{S(N)}$ parameters 
(characteristic kinetic energies of the condensate in the two layers), 
introduced here, depend on the magnetic field and momentum $q_\mathrm{cm}$:
\begin{align}
        E_S &= \frac{E_{\mathrm{Th},S} d_S}{2}   \int_0^{d_S} q_s^2(x) dx,
        \label{eq:ES_appering}
        \\
        E_N  &= \frac{E_{\mathrm{Th},N} d_N}{2}   \int_{-d_N}^{0} q_s^2(x) dx,
        \label{eq:EN_appering}
\end{align}
where
\begin{equation} \label{eq:Thouless_def}
    E_{\mathrm{Th},S(N)} = D_{S(N)}/d_{S(N)}^2
\end{equation}
is the Thouless energy
(inverse diffusion time)
for each layer.
We have also introduced the characteristic ``escape times''
\footnote{The $\tau_{S(N)}$ parameters defined in Eq.\ \eqref{eq:tau_appearance}, can be estimated as $\tau_{S(N)} \sim \mathcal{T}^{-1}d_{S(N)}/v_{S(N)}$,
where $\mathcal{T}$ is the transparency of the interface and $v_{S(N)}$ are the Fermi velocities in the layers. Note that $\tau_{S(N)}$ do not depend on the mean free paths $l_{S(N)}$ in the layers and contain the ballistic times $d_{S(N)}/v_{S(N)}$ of passage across the layers. 
If the transverse resistance of the bilayer is mainly determined by the interface, $r_B \gg r_{S(N)}$ [equivalently, $\mathcal{T} \ll l_{S(N)}/d_{S(N)}$], 
the $\tau_{S(N)}$ parameters have a meaning of escape times from the corresponding layers \cite{Fominov2002}.
In the opposite limit of an ideal interface ($\mathcal{T}\sim 1$), we have $\tau_{S(N)} \sim d_{S(N)}/v_{S(N)}$, i.e., they have a form of \emph{ballistic} escape times (even though we consider a diffusive system). Having made the above remarks, we will refer to $\tau_{S(N)}$  as the escape times.}:
\begin{equation}
    \tau_{S(N)} = \frac{2 r_B}{r_{S(N)}} E_{\mathrm{Th},S(N)}^{-1} ,\quad
    \tau^{-1} = \tau_S^{-1} + \tau_N^{-1}.
    \label{eq:tau_appearance}
\end{equation}
The escape times $\tau_{S(N)}$ and $\tau$ contain information about the materials of the layers and the resistance of the interface: $\tau_{S(N)},\tau \propto r_B$.
Below, we characterize the interface resistance in terms of $\tau$, which can thus be considered as a measure of $r_B$.

Thus, we have expressed the densities at the interface $n_{S(N)0}$ and the $d$ corrections $\delta n_{S(N),d}$ in terms of the parameters at the interface $\theta_{S(N)0}$, which are determined from Eqs.\ \eqref{eq:main_eq_neproz1} and \eqref{eq:main_eq_neproz}. Corrections $\delta n_{S(N),d}$ are necessary because they are responsible for the SDE.

However, Eqs.\ \eqref{eq:main_eq_neproz1} and \eqref{eq:main_eq_neproz} cannot be solved analytically for an arbitrary interface resistance, i.e., arbitrary $\tau_{S(N)}$.
We will therefore examine the limiting cases of weakly ($\tau \to 0$, see Sec.\ \ref{sec:Weak_case}) and strongly ($\tau \to \infty$, see Sec.\ \ref{sec:Strong_case}) resistive interface.
As discussed in Secs.~\ref{sec:Weak_conditions} and \ref{sec:Strong_conditions} below, the comparison scale for $\tau$ is the coherence time $\Delta^{-1}$; the limits $\tau \to 0$ and $\tau \to \infty$ thus correspond to $\tau\Delta\ll 1$ and $\tau\Delta\gg 1$, respectively.
In these cases, the parameters $\theta_{S(N)0}$ and the densities $n_{S(N)0}$ at the interface can be expanded with respect to an additional small parameter, either $\tau$ or $\tau^{-1}$ for the case of weak or strong resistance, respectively:
\begin{equation} \label{eq:tau_decompose}
    \theta_{S(N)0} = \theta_{S(N)}^{(0)} + \delta \theta_{S(N),\tau}, \quad n_{S(N)0} = n_{S(N)}^{(0)}+ \delta n_{S(N),\tau}.
\end{equation}
Here, $\theta_{S(N)}^{(0)}$ and $n_{S(N)}^{(0)}$ are the spectral angles and densities at the interface in the limit $\tau = 0$ or $\infty$ (depending on the limiting case under consideration), while $\delta \theta_{S(N),\tau}$ and $\delta n_{S(N),\tau}$ are corrections 
related to small but finite $\tau$ or $\tau^{-1}$, respectively. In both cases, we call them $\tau$ corrections;
they are related to the interface jump of the solution.

The densities $n_{S(N)}^{(0)}$ and $\tau$ corrections $\delta n_{S(N),\tau}$ can be obtained by expanding Eq.\ \eqref{eq:plotnost} with $\theta_{S(N)} = \theta_{S(N)0}$ from Eq.\ \eqref{eq:tau_decompose}. 
The superfluid density in the zeroth order and the corrections then take the following form:
\begin{gather}
    n_{S(N)}^{(0)} = \frac{2 \pi m \sigma_{S(N)} T }{e^2} \sum_{\omega_n>0} \sin^2 \theta_{S(N)}^{(0)}, \label{eq:density_zero_order_def}\\
    \delta n_{S(N),\tau} = \frac{2 \pi m \sigma_{S(N)} T }{e^2} \sum_{\omega_n>0} \delta \theta_{S(N),\tau}\sin2 \theta_{S(N)}^{(0)},
    \label{eq:tau_corection_density}\\
    \delta n_{S(N),d} = \frac{2 \pi m \sigma_{S(N)} T }{e^2} \sum_{\omega_n>0} \delta \theta_{S(N),d} \sin2  \theta_{S(N)}^{(0)}.
    \label{eq:corection_density}
\end{gather}

The density corrections from Eqs.\ \eqref{eq:tau_corection_density} and \eqref{eq:corection_density} are needed to expand the current in Eq.\ \eqref{eq:current_general}. These corrections contribute to the average superfluid density and the center-of-mass coordinate:
\begin{equation}
    \bar n = \bar n^{(0)} + \delta \bar n, \quad x_\mathrm{cm} = x_\mathrm{cm}^{(0)}+\delta x_\mathrm{cm}.
    \label{eq:diode_start}
\end{equation}
Here, $\bar n^{(0)}$ and  $x_\mathrm{cm}^{(0)}$ correspond to the zeroth approximation (with respect to $\tau$ and $d$ inhomogeneities),
\begin{equation}
    \bar n^{(0)} = \frac{n_{S}^{(0)} d_S + n_{N}^{(0)} d_N}{d},\quad
    x_{\mathrm{cm}}^{(0)} = \frac{n_{S}^{(0)} d_S^2 - n_{N}^{(0)} d_N^2}{2\bar n^{(0)} d},
    \label{eq:center_of_mass}
\end{equation}
while the first-order corrections are given by
\begin{align}
    &\delta \bar n = \frac{1}{d} \bigl[ (\delta n_{S,\tau} + 2 \delta n_{S,d}/3) d_S + (\delta n_{N,\tau} - 2\delta n_{N,d}/3) d_N \bigr], 
    \label{eq:average_density_correction} \\
\label{eq:center_of_mass_correction}
    &\delta x_\mathrm{cm} = \frac{1}{2 \bar n^{(0)} d } \bigl[ (\delta n_{S,\tau} + 5\delta n_{S,d}/6) d_S^2 \notag \\
    &\hphantom{\delta x_\mathrm{cm} =}
    - (\delta n_{N,\tau} - 5\delta n_{N,d}/6) d_N^2 \bigr] - x_{\mathrm{cm}}^{(0)}\frac{\delta \bar n}{\bar n ^{(0)}}.
\end{align}

In the leading order, the density "knows" only about the center-of-mass momentum in the zeroth order: $\bar n^{(0)} = \bar n^{(0)} (q^{(0)}_\mathrm{cm})$, with $q^{(0)}_\mathrm{cm} = q_0 - 2 e B x_\mathrm{cm}^{(0)}$. Defining the deviation of the center-of-mass momentum from the zeroth order according to
\begin{gather}
    q_\mathrm{cm} = q_\mathrm{cm}^{(0)} - \delta q_\mathrm{cm},
    \\
    \delta q_\mathrm{cm}(q_\mathrm{cm},B) = 2 e B \delta x_\mathrm{cm}(q_\mathrm{cm},B), 
\end{gather}
we can write Eq.\ \eqref{eq:diode_start} in a detailed form:
\begin{equation}
    \bar{n} (q_\mathrm{cm},B) = \bar{n}^{(0)}(q_\mathrm{cm}^{(0)},B) + \delta \bar{n}(q_\mathrm{cm},B).
\end{equation}
Together with Eq.\ \eqref{eq:current_general}, this gives the expansion for the current:
\begin{multline}
    \frac{I(q_{\mathrm{cm}},B)}{I_\mathrm{ch}} = -\frac{1}{\bar n^{(0)}_0}\Big[\bar n^{(0)}\bigl (q_\mathrm{cm} + 
    \delta q_\mathrm{cm}(q_\mathrm{cm},B),B \bigr)  \\
    + \delta \bar n(q_\mathrm{cm},B) \Big] \frac{q_\mathrm{cm}}{q_c}.
    \label{eq:current_main}
\end{multline}
Here, we have introduced the characteristic value of the superfluid density $\bar n_0^{(0)}$, which is defined as the zero-temperature value of $\bar n^{(0)}(0,0)$. We have also introduced $q_c$, the critical center-of-mass momentum, in the sense of $\bar n^{(0)}(q_c,0) = 0$. The explicit expression for $q_c$ will be given below [see Eqs.\ \eqref{eq:critical_q_B} and \eqref{eq:critical_q_B_2}]. The current is normalized to the characteristic value
\begin{equation} \label{eq:Ich}
    I_\mathrm{ch} = (e w d/2m) \bar n^{(0)}_0 q_c .
\end{equation} 

The SDE arises if the expression in the square brackets in Eq.\ \eqref{eq:current_main} is not an even function of the momentum [see Eq.\ \eqref{eq:diode_criteria}]. As we will see below, in the zeroth order there is no diode effect. This implies that $\bar n^{(0)}$ is an even function of both its arguments, hence we need to consider the $\delta \bar n$ and $\delta q_\mathrm{cm}$ corrections. Therefore, we will only be interested in the even (with respect to $q_\mathrm{cm}$) part of the $\delta q_\mathrm{cm}(q_\mathrm{cm},B)$ function and the odd part of the $\delta \bar n(q_\mathrm{cm},B)$ function. The asymmetry can result from any of the terms in the square brackets.

\section{Weakly nonideal interface in thin bilayers}
\label{sec:Weak_case}

Now, we apply the general consideration of the previous section to the case of a weakly nonideal interface, $\theta_{S0} \approx \theta_{N0}$.
Deviations from the zeroth-order solutions are treated perturbatively. 
Two different types of corrections arise: those due to the finite thickness of the bilayer $d\neq 0$ ($d$ corrections) and those due to the finite interface resistance, i.e., $\tau \neq 0$ ($\tau$ corrections).
For each type, we keep only the leading-order corrections.
We schematically demonstrate the two types of corrections in Fig.~\ref{fig:squares1}.

\subsection{Zeroth order}
\label{sec:ideal}

The zeroth order corresponds to the ideal interface and homogeneity of the spectral angle across the whole bilayer.

In the case of a fully transparent interface ($r_B = 0$, hence $\tau = 0$), the spectral angles are equal at the interface: $\theta_{S0} = \theta_{N0} = \theta$, hence $\theta_j =0$. In this case, the densities in the zeroth order $n_{S(N)}^{(0)}$ [see Eq.\ \eqref{eq:density_zero_order_def}] differ only due to the difference of the conductivities: $n_{S}^{(0)}/n_{N}^{(0)} = \sigma_S/\sigma_N$. The center-of-mass coordinate [see Eq.\ \eqref{eq:center_of_mass}] is then given by
\begin{equation}
    x_{\mathrm{cm}}^{(0)} = \frac{\sigma_S d_S^2 -\sigma_N d_N^2}{2(\sigma_S d_S +\sigma_N d_N)} = \mathrm{const}_{q_\mathrm{cm},B}.
    \label{eq:center_of_mass_1}
\end{equation}

Equations \eqref{eq:main_eq_neproz1} and \eqref{eq:main_eq_neproz} determining the interface values $\theta_{S(N)0}$ then reduce to a single equation for the spectral angle $\theta$:
\begin{equation}
    -\frac{\omega_n}{\cos \theta(\omega_n)} +  \frac{E_g}{\sin \theta(\omega_n)}  = \Gamma.
    \label{eq:main_equation}
\end{equation}
In this limit, the system behaves as a homogeneous superconductor with effective order parameter $E_g$. Therefore, the equation includes only some average values $E_g$ and $\Gamma$ that characterize the bilayer as a whole:
\begin{gather}
    E_g = \langle \Delta \rangle, 
    \label{eq; def_minigap}
    \\
    \Gamma = \langle E \rangle =\frac{\langle D \rangle }{2} \left[q_\mathrm{cm}^2+\frac{1}{3}(e B d_{\mathrm{eff}})^2
        \right]. \label{eq:Gamma_def}
\end{gather}
Here, we have introduced the effective thickness $d_\mathrm{eff}$ and the bracket notation $\langle\dots\rangle$ for averaging between values in the two layers [see Eq.\ \eqref{eq:bracket_averaging_def} below].

Equation \eqref{eq:main_equation} has the same form as in the Abrikosov-Gor'kov (AG) theory \cite{Abrikosov1960} for a superconductor containing paramagnetic impurities. In this theory, the $\Gamma$ parameter is the magnetic scattering rate. By analogy with this theory, we call $\Gamma$ the pair-breaking parameter.

The bracket $\langle\dots\rangle$ denotes averaging over layers with a weight $\sigma d/D$
(see also comment
\footnote{Using the relation $\sigma =2 e^2  \nu D$ between the conductivity and the diffusion constant, we can rewrite Eq.\ \eqref{eq:bracket_averaging_def} in terms of the total (volume-integrated) density of states $\nu_{S(N)} V_{S(N)}$ or in terms of the level spacings $\delta_{S(N)}$:\\
$\displaystyle\langle F\rangle = \frac{F_{S} \nu_{S} V_{S} + F_{N} \nu_{N} V_{N}}{\nu_{S} V_{S} + \nu_{N} V_{N}} = \frac{F_{S} \delta_{S}^{-1} + F_{N} \delta_{N}^{-1}}{\delta_{S}^{-1} + \delta_{N}^{-1}}$.
}):
\begin{equation}
    \langle F\rangle = \frac{F_S \sigma_S d_S/D_S +  F_N \sigma_N d_N/D_N}{\sigma_S d_S/D_S + \sigma_N d_N/D_N} = \frac{ F_S \tau_S +  F_N \tau_N}{\tau_S + \tau_N}
    \label{eq:bracket_averaging_def}
\end{equation}
(the last expression is physically transparent and has a well-defined limit at $r_B\to 0$ since the ratio $\tau_S/\tau_N$ remains finite in this case).
This averaging makes sense only when the inhomogeneities in each layer are weak, and the value of $F$ is constant in each layer.
This is exactly the case in the zeroth order with respect to the inhomogeneities, which we discuss now. Note that averaging of the order parameter should be understood as averaging between $\Delta_S(x) = \Delta$ and $\Delta_N(x) =0$.
At $\Gamma=0$, the quantity $E_g =\tau_S \Delta /(\tau_S + \tau_N)$ defined by Eq.\ \eqref{eq; def_minigap} yields the spectral gap in the density of states in the bilayer \cite{Abrikosov1960, McMillan1968, Golubov1994, Golubov1996, Fominov2001.PhysRevB.63.094518}.

The pair-breaking parameter $\Gamma$ in Eq.\ \eqref{eq:Gamma_def} can be interpreted as the average kinetic energy of the condensate (specifically, an average of the two kinetic energies). The first term, $\langle D \rangle q_\mathrm{cm}^2/ 2$, is the kinetic energy of the center-of-mass motion. The second term, $\langle D \rangle(e B d_{\mathrm{eff}})^2 /6$, should be understood as the kinetic energy of rotation relative to the center of mass. The effective thickness entering this term is the mean-square thickness, similar to that which enters the moment of inertia:
\begin{multline}
        d_{\mathrm{eff}}^2 =\frac{12}{\bar n^{(0)} d} \int n(x) \left(x-x_\mathrm{cm}^{(0)}\right)^2 dx\\
        =  \frac{\sigma_S d_S^3 + \sigma_N d_N^3}{\sigma_S d_S + \sigma_N d_N} + \frac{3\sigma_S\sigma_N d_S d_N(d_S + d_N)^2}{(\sigma_S d_S + \sigma_N d_N)^2}.
\end{multline}
The value of $d_{\mathrm{eff}}$ is the effective thickness of the bilayer from the point of view of the in-plane critical magnetic field.
This critical field at zero temperature is then expressed in the standard BCS form:
\begin{equation}
    B_{c0}  = \sqrt{3} \Phi_0 / \pi \xi_0 d_\mathrm{eff},
    \label{eq:critical_B_zero_temperature}
\end{equation}
with the effective coherence length $\xi_0$ defined as
\begin{equation}
    \xi_0 = \sqrt{\langle D \rangle/E_{g0}}. 
\end{equation}
Here, $E_{g0} = E_g(T=0, \Gamma=0)$ is the spectral gap at zero temperature.

It is convenient to express the pair-breaking parameter in terms of the temperature-dependent critical values $\Gamma_c$, $q_c$, and $B_c$:
\begin{equation}
    \Gamma = \Gamma_c(T) \left(\frac{q_\mathrm{cm}^2}{q_c^2(T)}+\frac{B^2}{B_c^2(T)}\right).
    \label{eq:pair-breaking_criticals}
\end{equation}
Here, $\Gamma_c(T)$ is the critical pair-breaking parameter, which corresponds to the phase transition to a normal state. The critical values of the momentum and magnetic field can be expressed in terms of the pair-breaking parameter:
\begin{equation}
    q_c(T) = \sqrt{\frac{2 \Gamma_c(T)}{\langle D\rangle}},\quad
    B_c(T) = \frac{\sqrt{3} \Phi_0 q_c(T)}{\pi d_{\mathrm{eff}}} .
    \label{eq:critical_q_B}
\end{equation}
[Note that $B_c(0) =B_{c0}$, see Eq.\ \eqref{eq:critical_B_zero_temperature}.] The critical momentum $q_c$ corresponds to the point at which the kinetic energy of the condensate equals $\Gamma_c$.

In the case of a weakly nonideal interface, superconductivity exists at temperatures $T < T_{c0}$, where $T_{c0}$, defined by Eq.\ \eqref{eq:Tc0}, is the critical temperature of the bilayer in the absence of both the current and magnetic field.

Thus, in the zeroth order
(with respect to the $d$ and $\tau$ corrections), 
the system ``feels'' the momentum $q_\mathrm{cm}$ and the magnetic field $B$ exclusively through the pair-breaking parameter: $E_g(q_\mathrm{cm},B) = E_g(\Gamma)$, $\bar n^{(0)}(q_\mathrm{cm},B) = \bar n^{(0)}(\Gamma)$. At the same time, $\Gamma$ depends on $q_\mathrm{cm}$ and $B$ in an even manner: $\Gamma(q_\mathrm{cm},B)=\Gamma(-q_\mathrm{cm},B) = \Gamma(q_\mathrm{cm},-B)$. This means that in the zeroth order, no symmetry breaking occurs, and consequently, the SDE is absent.

Alternatively, the absence of the SDE in the zeroth order can be related to the behavior of the superfluid density profile $n(x)$. Although nonconstant in space, this profile has a trivial steplike form that remains fixed by the relation $n_{S}^{(0)}/n_{N}^{(0)} = \sigma_S/\sigma_N$.
This implies that while the overall amplitude of $n(x)$ may change, the \emph{profile} of this functions does not vary with $I$ and $B$, and this forbids the SDE.

In order to discuss the SDE in higher orders, we introduce the dimensionless parameter of asymmetry between the layers
\begin{equation} \label{eq:mu}
    \mu =\frac{\Delta-E_g}{\Delta} =\frac{\sigma_N d_N/D_N}{\sigma_N d_N/D_N + \sigma_S d_S/D_S} = \frac{\tau_N}{\tau_N + \tau_S },
\end{equation}
which can be interpreted as the fraction of time spent by quasiparticles in the normal layer.
In terms of $\mu$, the averaging operation defined earlier can be rewritten in a more compact form:
\begin{equation}
    \langle F\rangle = (1-\mu)F_S + \mu F_N .
\end{equation}
The absence of the SDE in the zeroth order is due to the fact that $\mu$ enters the description only through averages of this form, which yields an effective medium with averaged parameters. As we will see below, in higher orders, the asymmetry parameter $\mu$ will also enter the theory by itself, leading to the SDE.

\subsection{\texorpdfstring{$d$ corrections}{d corrections}}
\label{sec:d_corr}

\begin{figure}[t]
 \includegraphics[width=\columnwidth]{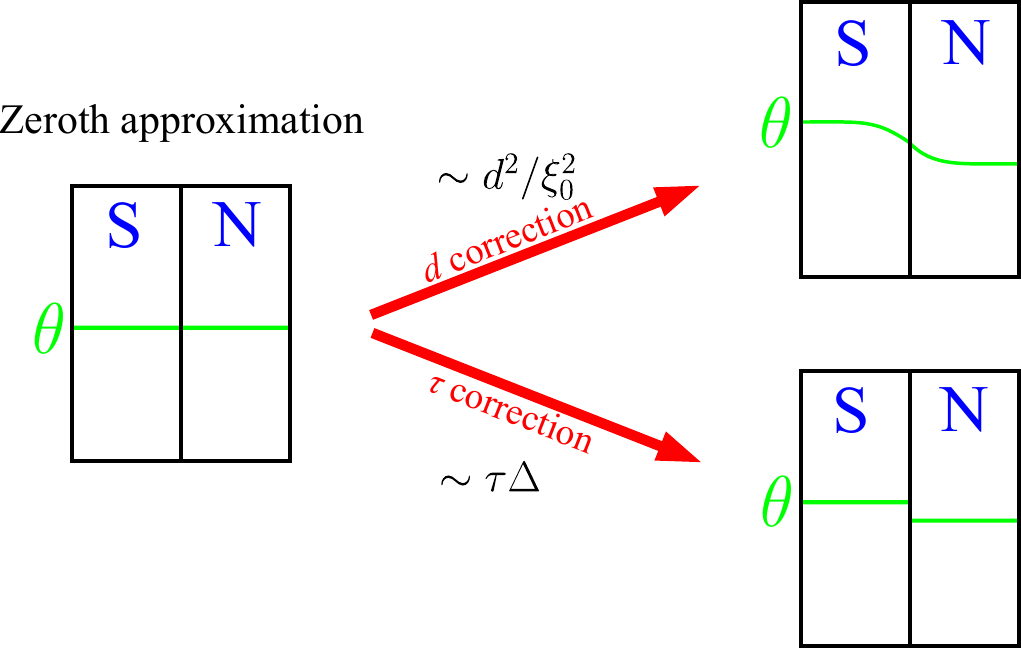}
 \caption{Schematic representation of the leading-order $d$ and $\tau$ corrections to the spectral angle in the limit of a weakly nonideal interface in a thin bilayer. 
 The $d$ correction (intralayer correction) is related to weak inhomogeneity within each layer, while the $\tau$ correction is related to the interface jump of the solution (weak interfacial inhomogeneity).
 The orders of magnitude of the corrections (shown near the red arrows) follow from the analysis in Sec.\ \ref{sec:Weak_case}.} 
 \label{fig:squares1}
\end{figure}

The $d$ corrections from Eq.\ \eqref{eq:soultions_neodnor} in the case of ideal interface are as follows: 
\begin{align}
        \delta \theta_{S,d} &= \frac{ \sin \theta \cos \theta}{E_{\mathrm{Th},S}} 
        \left(
        -\frac{\omega_n}{\cos \theta}  - E_S + \frac{\Delta}{\sin \theta}
         \right),\label{eq:solutions1}\\
        \delta \theta_{N,d} &= \frac{ \sin \theta \cos \theta}{E_{\mathrm{Th},N}}
        \left(
        -\frac{\omega_n}{\cos \theta} - E_N
        \right).
        \label{eq:solutions}
\end{align}
These corrections are small (hence, valid) as long as $\omega_n \ll E_g \xi_0^2/d^2$. At the same time, this range provides the dominant contribution to the sum for the density corrections in Eq.\ \eqref{eq:corection_density}. A schematic representation of this correction is shown in Fig.~\ref{fig:squares1} (upper arrow).

The series in Eq.\ \eqref{eq:corection_density} then produces a logarithmic result:
\begin{equation}
    \frac{\delta n_{S(N),d}}{\bar n_0^{(0)}} = \frac{4}{\pi} \mu \;\frac{E_g^2}{E_{g0}^2} \frac{ d_{S(N)}d}{\xi_0^2}\; \ln \frac{\xi_0^2}{d^2}.
    \label{eq:density_correction_1}
\end{equation}
Here, $\bar n_0^{(0)} = \bar n^{(0)}(T=0,\Gamma=0)$ [see Eq.\ \eqref{eq:center_of_mass} with $n^{(0)}_{S(N)}(0,0) = \pi m E_{g0} \sigma_{S(N)}/e^2$] is the average superfluid density at zero temperature in the absence of a magnetic field and current in the zeroth order with respect to inhomogeneity.

The main nonreciprocal contribution to the current is then given by Eq.\ \eqref{eq:current_main} with
    \begin{equation}
        \delta q_\mathrm{cm} = k q_c \frac{\bar n_0^{(0)}}{\bar n^{(0)}(\Gamma)} \left(\frac{E_g(\Gamma)}{E_{g0}}\right)^2  \frac{B}{B_c} ,
        \label{eq; delta_q1}
    \end{equation}
and without $\delta \bar n(\Gamma)$, because the latter is an even function of the momentum (not producing the SDE). The inhomogeneity parameter $k$ introduced here characterizes the strength of the SDE. As we will see below, generally, it includes both $d$ and $\tau$ contributions ($k = k_d + k_\tau$). 

What we have obtained up to now is the $d$ contribution ($k = k_d$), which is small due to the smallness of the $d$ inhomogeneities:
\begin{multline}
    k_{d} = 2\sqrt{3} \frac{\delta x_\mathrm{cm}(0,0)}{d_\mathrm{eff}}
    \\
    =\mu \frac{5-(d_S-d_N)(d_S-d_N + 32 L_{\mathrm{cm}})}{2 \pi \sqrt{3}d_{\mathrm{eff}}d} \left(\frac{d^2}{\xi_0^2}\right) \ln \frac{\xi_0^2}{d^2} .
    \label{eq:k_prozr}
\end{multline}
Here, $L_\mathrm{cm} =x_\mathrm{cm}^{(0)}- (d_S-d_N)/2$ is the displacement of the center of mass relative to the geometric center of the bilayer. The effective magnetic thickness can also be expressed in terms of these quantities:
\begin{equation}
    (d_{\mathrm{eff}}/d)^2 = 1 - 4 L_\mathrm{cm}(d_S-d_N+ 3 L_\mathrm{cm})/d^2.
    \label{eq:d_eff_new}
\end{equation}
The inhomogeneity parameter can be estimated as follows:
\begin{equation} \label{eq:kd}
    k_{d} \sim \frac{d^2}{\xi_0^2} \ln \frac{\xi_0^2}{d^2} \ll 1.
\end{equation}
The condition of smallness, $k_d \ll 1$, implies that $d \ll \xi_0$, which can be rewritten as $\Delta \ll \langle D\rangle / d^2$. The latter expression has the meaning of the effective Thouless energy of the whole bilayer.

The numerator in Eq.\ \eqref{eq:k_prozr} is always positive. This is due to the fact that at fixed $(d_S-d_N)$, the value of $L_\mathrm{cm}$ can only vary within the range $-d_S/2<L_\mathrm{cm}<d_N/2$. 

Equations \eqref{eq:current_main} and \eqref{eq; delta_q1} demonstrate the SDE due to transverse superconducting inhomogeneities ($\nabla n$ in Fig.~\ref{fig:bilayr}). The variation of the $n(x)$ profile is caused by nonzero $\delta q_\mathrm{cm}$ and nontrivial $\delta x_\mathrm{cm} \neq \mathrm{const}_{q_\mathrm{cm},B}$, which break the symmetry  \eqref{eq:diode_criteria} and thus produce the SDE. 

The strength of the SDE due to the $d$ inhomogeneities is characterized by $k_d$ from Eq.\ \eqref{eq:k_prozr}. In the absence of inhomogeneities, $k_d =0$. For example, inhomogeneities vanish in the absence of the N layer, $d_N\to 0$, or in the case of poorly conducting N layer, $\sigma_N \to 0$. Inhomogeneities are also suppressed with decreasing the thickness of the bilayer, $d \to 0$. At the same time, decreasing the thickness or conductance of the S layer ($d_S \to 0$ or $\sigma_S \to 0$) does not lead to vanishing of superconducting inhomogeneities in the bilayer as a whole, and $k_d$ remains finite. This is so because inhomogeneities in this case vanish only in the S layer, while the induced superconductivity in the N layer (of fixed thickness) remains inhomogeneous. [Still, the SDE vanishes in this limit because $\delta q_{\mathrm{cm}}\to 0$ due to $q_c \to 0$ in Eq.\ \eqref{eq; delta_q1}.] 

The $k_d$ parameter is a nontrivial function of the bilayer's parameters. Analyzing Eq.\ \eqref{eq:k_prozr}, we conclude that in the thin-bilayer limit, the SDE can be maximized by increasing the total thickness of the bilayer $d$ while keeping moderate ratios $d_S /d_N \sim 1$ and $\sigma_S/ \sigma_N \sim  1$.

Explicit analytical expressions (as a function of $q_\mathrm{cm}$ and $B$) for $\delta q_\mathrm{cm}$ in Eq.\ \eqref{eq; delta_q1} and the corresponding current in Eq.\ \eqref{eq:current_main} can be obtained only in the limiting cases of the AG theory, at zero temperature ($T=0$) and near the phase transition ($\Gamma \to \Gamma_c$). These cases will be discussed later in Sec.\ \ref{sec:AGlimits}.

The $d$ corrections for $\theta$ inevitably imply the inhomogeneity of the order parameter, i.e., the presence of $\Delta_1(x)$. At the same time, our consideration has demonstrated that the SDE appears already on the background of a homogeneous order parameter $\Delta$. The correction $\Delta_1(x)$ can therefore be disregarded since it would only lead to small corrections to our results.

\subsection{\texorpdfstring{$\tau$ corrections}{tau corrections}}
\label{sec:int_corr}

Next, we take into account the corrections related to the finite resistance (nonideal transparency) of the interface, $\tau \neq 0$. In the weakly nonideal limit, the jump of the spectral angle is small, $\delta \theta_{S(N),\tau} = \pm\theta_j/2\ll\theta$. Here, the $\theta$ parameter is determined from the zeroth-order Eq.\ \eqref{eq:main_equation}. The corrections can then be found by expanding either Eq.\ \eqref{eq:main_eq_neproz1} or \eqref{eq:main_eq_neproz}:
\begin{multline}
     \theta_j = \tau_S \sin \theta \cos \theta \left(-
    \frac{\omega_n}{\cos \theta} - E_S + \frac{\Delta}{\sin \theta}
    \right)  \\
    =  -\tau_N \sin \theta \cos \theta \left(
    -\frac{\omega_n}{\cos \theta} - E_N
    \right).
    \label{eq:small_jump}
\end{multline}
The $\delta \theta_{S(N),\tau}$ corrections are small as long as $\omega_n \ll 1/\tau_N$. A schematic representation of these corrections is shown in Fig.~\ref{fig:squares1} (lower arrow).

The series in Eq.\ \eqref{eq:tau_corection_density} then produces a logarithmic result
\begin{equation}
    \frac{\delta n_{S(N),\tau}}{\bar n^{(0)}_0} = \pm \frac{1}{\pi} \mu\frac{D_{S(N)}d}{\langle D\rangle d_{S(N)}} \frac{E_{g}^2 \tau_N \tau_{S(N)}}{E_{g0}\tau} \ln \frac{1}{E_{g0}\tau_N}.
    \label{eq:density_tau_corr_weak}
\end{equation}
[The calculation is similar to the case of the $d$ corrections in Eq.\ \eqref{eq:density_correction_1}.]

As a result, the SDE is determined by Eqs.\ \eqref{eq:current_main} and \eqref{eq; delta_q1} with the following inhomogeneity parameter:
\begin{gather}
    k = k_d + k_{\tau}, 
    \label{eq:full_k1}\\
        k_{\tau} = \mu\frac{8\sqrt{3}}{\pi}\frac{D_S D_N}{\langle D \rangle^2} \frac{d}{d_{\mathrm{eff}}}E_{g0}\tau \ln \frac{1}{E_{g0}\tau_N},
    \label{eq:full_k}
\end{gather}
and without $\delta \bar n$, because the latter is an even function of the momentum (not producing the SDE). 

The $k$ parameter in Eq.\ \eqref{eq:full_k1} contains contributions of both the $d$ and $\tau$ inhomogeneities and determines the total strength of the SDE in our system.
The total parameter $k$ can be interpreted as the normalized center-of-mass shift: $k \sim \delta x_\mathrm{cm}/d_\mathrm{eff}$.
Since the $\tau$ contribution makes the total inhomogeneity parameter $k$ larger, it enhances the SDE. Equation \eqref{eq:full_k} thus demonstrates that the SDE monotonically grows with increasing $\tau$ in the limit of the weakly nonideal interface. 

According to Eq.\ \eqref{eq:full_k1} the SDE appears even if $k_d=0$. Despite the fact that the superfluid density $n(x)$ in this case is a steplike function, the $\tau$ corrections make the ratio between the densities $n_{S0}/n_{N0}$ sensitive to the current and magnetic field, which implies variation of the profile [see Eqs.\ \eqref{eq:tau_decompose} and \eqref{eq:density_tau_corr_weak}].

\subsection{Conditions of applicability }
\label{sec:Weak_conditions}

Now we discuss the applicability conditions of the above solutions. Two types of conditions should be discussed in this respect. The first one is the condition of weakly nonideal interface, $\theta_j \ll \theta$. The second one is the condition of weak intralayer inhomogeneity, $\delta \theta_{S(N),d} \ll 1$. Both the conditions should be considered at characteristic frequencies $\omega_n \gtrsim E_g$ providing the main contribution to the sums in Eqs.\ \eqref{eq:density_zero_order_def}-\eqref{eq:corection_density} for the superfluid densities .

The requirement of weak interface resistance, $\theta_j \ll \theta$, according to Eq.\ \eqref{eq:small_jump} is satisfied at
\begin{equation}
     \tau_{S(N)}\Delta \ll 1. \label{eq:appl_weak_intermediate}
\end{equation}
Thus, the condition on the interface resistance reduces to a relationship between the interlayer transfer time and the coherence time $\Delta^{-1}$ (the time it takes for a quasiparticle to travel a coherence length). This means that quasiparticles leave their layer before decoherence sets in.

Information about the interface is included in Eq.\ \eqref{eq:appl_weak_intermediate} not only directly due to $\tau_{S(N)}$, but also due to  $\Delta$, because its maximum value $\Delta_0 = \Delta(T=0,\Gamma =0)$ depends on the escape times $\tau_{S(N)}$ \cite{Fominov2001.PhysRevB.63.094518}:
\begin{equation}
    \frac{\Delta_0}{\Delta_{S0}} = \frac{\tau_N}{\tau}\left(\frac{\Delta_{S0}}{2\omega_D} \sqrt{1+ (\omega_D\tau)^2}\right)^{\tau_N/\tau_S}.
    \label{eq:max_Delta_0}
\end{equation}
Here, $\omega_D$ is the Debye frequency, $\Delta_{S0} = T_{cS}\pi / e^\gamma$ is the order parameter of an isolated S layer at zero temperature in the absence of a magnetic field and current, and $\gamma \approx0.577$ is Euler's constant. 

With the help of condition $\Delta < \Delta_0$, we can simplify the condition in Eq.\ \eqref{eq:appl_weak_intermediate} as
\begin{equation}
     \tau\Delta \ll 1. \label{eq:appl_weak_11}
\end{equation}
This condition also ensures that $\delta n_{S(N),\tau} \ll n_{S(N),0}$.

The above conditions can be rewritten as $\tau_S \Delta \ll 1$ and $\tau_N E_g \ll 1$, which can be interpreted in the language of characteristic lengths. The distance $\sqrt{D_{S(N)}\tau_{S(N)}}$ traveled during time $\tau_{S(N)}$, is much smaller than the coherence length of the corresponding layer $\xi_{S(N)}$:
\begin{gather}
    \sqrt{D_{S(N)}\tau_{S(N)}}\ll \xi_{S(N)},
    \\
    \xi_S = \sqrt{D_S/\Delta}, \quad \xi_N = \sqrt{D_N/E_g}.
    \label{eq:coherence_length}
\end{gather}

The second requirement, $\delta \theta_d \ll 1$, according  to Eqs.\ \eqref{eq:solutions1} and \eqref{eq:solutions} is satisfied at
\begin{equation} \label{eq:appl_weak_2}
    d_{S} \ll \xi_{S}/\sqrt{\mu},\quad d_N \ll \xi_N.
\end{equation}
These conditions can also be written in an energy form with the help of the Thouless energies defined in Eq.\ \eqref{eq:Thouless_def}:
\begin{equation} \label{eq:condition_Thouless}
    \mu \Delta \ll E_{\mathrm{Th},S}, \quad (1-\mu) \Delta \ll E_{\mathrm{Th},N},
\end{equation}
which means that diffusion across each layer is faster than decoherence.

Summarizing, our consideration of the case of weak inhomogeneity and weak interface resistance is valid for the resistances specified by condition \eqref{eq:appl_weak_11} and for thin layers in the sense of condition \eqref{eq:appl_weak_2}.

\subsection{Comparison of contributions}
\label{sec:comparison}

We have obtained two contributions to the SDE: from the $d$ corrections (finite thickness of the bilayer) described by Eq.\ \eqref{eq:k_prozr} and from the $\tau$ corrections (finite interface resistance) described by Eq.\ \eqref{eq:full_k}. Comparing these two contributions, we get two regimes, in which one of the contributions dominates. 

Assuming the logarithms and diffusion constants $D_{S(N)}$ in Eqs.\ \eqref{eq:k_prozr} and \eqref{eq:full_k} of the same order, we find that condition $k_d \sim k_\tau$ would imply
\begin{equation}
     \tau \sim d^2/\langle D \rangle.
    \label{eq:perehod1}
\end{equation}
The latter expression has the meaning of the diffusion time across the bilayer. 

When the escape time \cite{Note2} is much less than the diffusion time, $\tau \ll d^2/\langle D \rangle$, the quasiparticles easily travel between the layers, and the $d$ contribution dominates, $k_d \gg k_\tau$. In this regime, the SDE is determined only by the geometry of the bilayer and is small as $\eta \sim \widetilde \eta \sim d^2/\xi_0^2$. 

If the escape time is much longer than the diffusion time, $d^2/\langle D \rangle \ll \tau \ll \Delta^{-1}$, then the quasiparticles spend most of their time in one of the layers, and the $\tau$ contribution to the SDE dominates, $k_\tau \gg k_d$. In this regime, the SDE increases with increasing the interface resistance, $\eta \sim \widetilde \eta \sim E_g \tau$.

\section{Strongly resistive interface in thin and moderately thick bilayers}
\label{sec:Strong_case}

\subsection{\texorpdfstring{$d$ and $\tau$ corrections}{d and tau corrections}}

Next, we consider the limiting case corresponding to a strongly resistive interface and weak superconducting proximity effect in the normal metal, such that $\theta_N \ll \theta_S$. In the zeroth-order approximation ($\tau \to \infty$), the S layer is effectively isolated, leading to $\theta_{N}^{(0)} = 0$ and $\theta_{S}^{(0)} = \mathrm{const}_x$. Within the perturbation theory, the $\tau$ corrections are incorporated into the interface values as follows:
\begin{equation}
    \theta_{S0} = \theta + \delta \theta_{S,\tau},\quad \theta_{N0} = \delta \theta_{N,\tau} \ll \theta,\quad
    \theta_j \approx\theta.
\end{equation}
The zeroth-order spectral angle $\theta$ is determined from the following equation:
\begin{equation}
    -\frac{\omega_n}{\cos \theta(\omega_n)} + \frac{\Delta}{\sin \theta(\omega_n)} = E_S,
    \label{eq:main_equation_2}
\end{equation}
which has the same AG form as Eq.\ \eqref{eq:main_equation}, but now it applies to an isolated S layer. The pair-breaking parameter here follows from Eq.\ \eqref{eq:ES_appering} with the center of mass located in the center of the S layer ($x_\mathrm{cm}^{(0)} = d_S/2$):
\begin{equation}
    E_S = \frac{D_S}{2} \left(q_\mathrm{cm}^2 + \frac{1}{3}(e B d_S)^2\right) = E_{Sc} \left(\frac{q_\mathrm{cm}^2}{q_c^2} + \frac{B^2}{B_c^2}\right). \label{eq:ES_def}
\end{equation}
The critical momentum and magnetic field are defined similarly to Eq.\ \eqref{eq:critical_q_B}, but for isolated S layer ($d_{\mathrm{eff}} \mapsto d_S$, $\Gamma\mapsto E_S$, $\langle D \rangle \mapsto D_S$):
\begin{equation}
    q_c(T) = \sqrt{\frac{2 E_{Sc}(T)}{D_S}},\quad
    B_c(T) = \frac{\sqrt{3} \Phi_0 q_c(T)}{\pi  d_S}.
    \label{eq:critical_q_B_2}
\end{equation}
In case of a strongly resistive interface, $T_{c0} \mapsto T_{cS}$. Therefore, $E_{Sc}(T)$ is determined by the equation having the same form as Eq.\ \eqref{eq:critical_pairbreaking} with $T_{c0} \mapsto T_{cS}$ and $\Gamma_c \mapsto E_{Sc}$.

\begin{figure}[t]
 \includegraphics[width=\columnwidth]{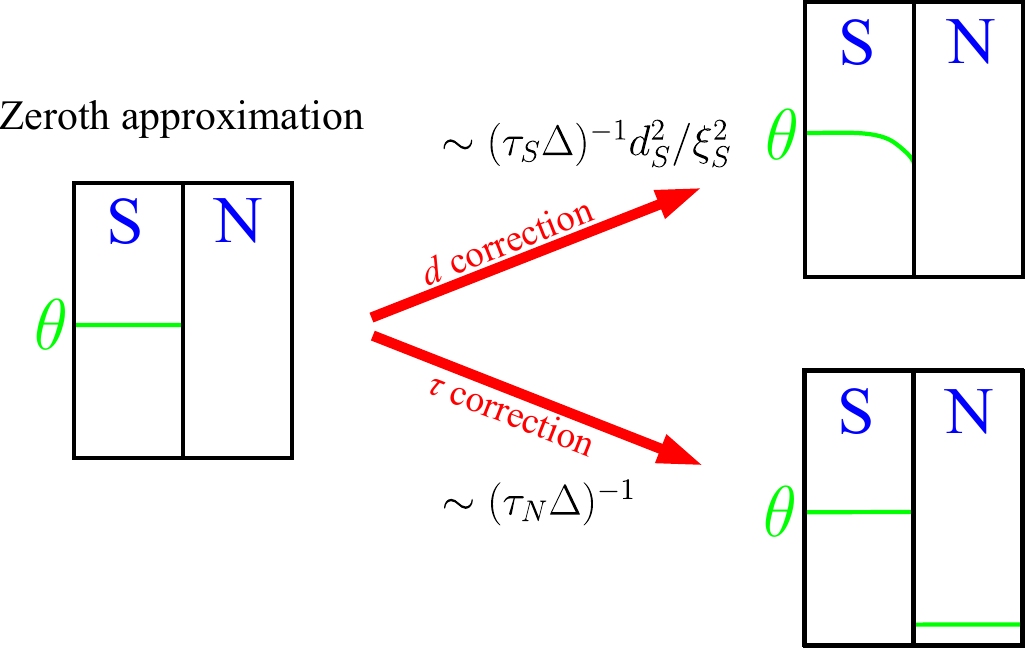}
 \caption{Schematic representation of the 
 leading-order $d$ and $\tau$ corrections to the spectral angle in the limit of a strongly resistive interface.
 The $d$ correction is related to weak inhomogeneity within each layer, while the $\tau$ correction is related to small deviation of the interface jump from the limit of completely decoupled layers. 
 The orders of magnitude of the corrections (shown near the red arrows) follow from the analysis in Sec.\ \ref{sec:Strong_case}.}
 \label{fig:squares2}
\end{figure}

The $\tau$ corrections are derived by expanding Eqs.\ \eqref{eq:main_eq_neproz1} and \eqref{eq:main_eq_neproz}, while the $d$ corrections are determined from Eq.\ \eqref{eq:soultions_neodnor} as before. As a result,
\begin{gather}
    \delta \theta_{S,\tau} = -\frac{ \sin \theta}{\tau_S(\omega_n \cos \theta +  E_S \cos 2 \theta +  \Delta\sin \theta)},
    \label{eq:nontr_corrections1}\\
    \delta\theta_{N,\tau} = \frac{ \sin \theta}{\tau_N(\omega_n + E_N)},\quad
    \delta \theta_{S(N),d} = \frac{\sin \theta}{\tau_{S(N)}E_{\mathrm{Th},S(N)} }.
    \label{eq:nontr_corrections}
\end{gather} 
A schematic representation of these corrections is shown in Fig.~\ref{fig:squares2}.

The $\tau$ correction in the S layer, $\delta \theta_{S,\tau}$, and both the $d$ corrections, $\delta \theta_{S(N),d}$, depend on the momentum $q_\mathrm{cm}$ and magnetic field $B$ only through $E_S$ and are even functions of both of them. At the same time, the $\tau$ correction in the N layer, $\delta \theta_{N,\tau}$, depends also on $E_N$  which takes the form
\begin{equation}
    E_N = \frac{D_N}{D_S} E_{Sc} \left[\frac{(q_\mathrm{cm}+e B d)^2}{q_c^2} + \frac{B^2}{B_c^2}\right].
\end{equation}
This is not an even function, $\delta\theta_{N,\tau}(-q_\mathrm{cm},B) \neq\delta\theta_{N,\tau}(q_\mathrm{cm},B) \neq \delta\theta_{N,\tau}(q_\mathrm{cm},-B)$, and this asymmetry produces a contribution to the SDE through $\delta \bar n$ [see Eq.\ \eqref{eq:current_main}].

The corrections given by Eqs.\ \eqref{eq:nontr_corrections1} and \eqref{eq:nontr_corrections}, together with the zeroth-order spectral angle from Eq.\ \eqref{eq:main_equation_2},  determine the corresponding density corrections $\delta n_{S(N),\tau}$ and $\delta n_{S(N),d}$ from Eqs.\ \eqref{eq:tau_corection_density} and  \eqref{eq:corection_density}, respectively. The Matsubara sums for these corrections cannot be calculated explicitly in the general case. However, they can be estimated as
\begin{align}
     \delta n_{S,\tau} &\sim \frac{n_S^{(0)}}{\tau_S \Delta}, &  \delta n_{N,\tau} &\sim \frac{\sigma_N}{\sigma_S}\frac{n_S^{(0)}}{( \tau_N \Delta)^{2}},\label{eq:konzetr_poradki1}\\
     \delta n_{S,d} &\sim  \frac{n_S^{(0)}}{\tau_S \Delta}\frac{d_S^2}{\xi_S^2},& \delta n_{N,d} &\sim \frac{\sigma_N}{\sigma_S} \frac{ n_S^{(0)}}{\tau_N\Delta} \frac{d_N^2}{\xi_N^2}.
    \label{eq:konzetr_poradki}
\end{align}
Here, $n_{S}^{(0)}$ is the zeroth-order superfluid density in the S layer defined by Eq.\ \eqref{eq:density_zero_order_def} with $\theta_S^{(0)} = \theta$ from Eq.\ \eqref{eq:main_equation_2}. The coherence lengths here are determined by Eq.\ \eqref{eq:coherence_length} with $E_g = 1/\tau_N$, where the latter quantity is the minigap in the density of states in the limit we are discussing now \cite{McMillan1968, Golubov1994, Golubov1996, Fominov2001.PhysRevB.63.094518}.
Note that $\xi_N \gg \xi_S$ in the limit of strongly resistive interface (which we discuss now). 

Although $\delta n_{S,\tau}$ may be the largest correction among the four given by Eqs.\ \eqref{eq:konzetr_poradki1} and \eqref{eq:konzetr_poradki}, it only modifies the total superfluid density within the S layer. Crucially, it is an even function of the momentum, $\delta n_{S,\tau} (-q_\mathrm{cm},B) = \delta n_{S,\tau} (q_\mathrm{cm},B) $, and thus does not contribute to the SDE. Therefore, we do not include it in $\delta q_\mathrm{cm}$ and $\delta \bar n$. 

The conditions of smallness of the $d$ corrections are satisfied in the limit of thin layers, $d_{S(N)} \ll \xi_{S(N)}$. However, these corrections remain small even for moderately thick layers, $d_{S(N)} \gtrsim \xi_{S(N)}$, provided the condition $\tau_{S(N)} \Delta\gg (d_{S(N)}/\xi_{S(N)})^2$ holds. This allows us to extend our analysis to the case of moderately thick layers with sufficiently resistive interface, characterized by large $\tau_{S(N)}$. Below, we consider these two cases.

\subsubsection{Limit of thin layers}

To begin with, we consider the case of thin layers, $d_{S(N)}\ll \xi_{S(N)}$. We also assume that the respective parameters of the layers are of the same order: $d_S\sim d_N\sim d$, $\sigma_S \sim \sigma_N$, and $\tau_S\sim \tau_N\sim \tau$. In this case, the primary corrections responsible for the SDE are $\delta n_{S,d} \sim \delta n_{S,\tau} (d_S/\xi_S)^2$ and $\delta n_{N,\tau}\sim \delta n_{S,\tau} (\tau\Delta)^{-1}$. The corresponding asymmetric corrections are given by
\begin{gather}
    \delta \bar n =  \delta n_{N,\tau} d_N/d,\label{eq:strong_res_density_corr}\\
    \delta q_\mathrm{cm} = 2 e B \left(\frac{\delta n_{S,d}}{12  n_S^{(0)}} d_S- \frac{\delta n_{N,\tau}}{n_{S}^{(0)}} \frac{d_N d}{d_S}\right).
    \label{eq:strong_res_center_of_mass}
\end{gather}
At the same time, the $\delta n_{N,d}$ correction is of a higher order of smallness, $\delta n_{N,d} \sim \delta n_{S,d} (\tau\Delta)^{-1} \sim \delta n_{N,\tau} (d_N/\xi_S)^2$, and can therefore be neglected. Note that $\delta n_{S,d}$ is omitted from Eq.\ \eqref{eq:strong_res_density_corr} because it is an even function of $q_\mathrm{cm}$ and does not contribute to the odd part of $\delta \bar n(q_\mathrm{cm})$.

The competition between the $\delta n_{S,d}$ and $\delta n_{N,\tau}$ contributions defines two regimes. In the limit of relatively small resistance, $\tau\Delta\ll\xi_S^2/d_S^2$, the $\delta n_{N,\tau}$ correction dominates, and, consequently, the SDE decreases as $\eta \sim \widetilde \eta \sim (\tau\Delta)^{-2}$. In the opposite limit $\tau\Delta\gg\xi_S^2/d_S^2$, the induced superconductivity is suppressed, the main role is played by inhomogeneity within the S layer, $\delta n_{S,d}$, and the SDE decreases as 
$\eta \sim \widetilde \eta \sim (d_S/\xi_S)^2 (\tau\Delta)^{-1}$.

In total, the current $I$ as a function of the momentum $q_\mathrm{cm}$ and magnetic field $B$ in the limit of thin layers is described  by Eq.\ \eqref{eq:current_main} with $\delta \bar n$ and $\delta q_\mathrm{cm}$ given by Eqs.\ \eqref{eq:strong_res_density_corr} and \eqref{eq:strong_res_center_of_mass}, respectively.

\subsubsection{Limit of moderately thick layers}
\label{sec:moderat_thick_l}

Now, we consider the case, in which at least one of the layers is moderately thick, $d_{S(N)}\sim \xi_S$ (recall that $\xi_S\ll\xi_N$ in the strongly resistive limit, which we consider now). In this regime, the $d$ correction in the S layer dominates over the corrections in the N layer, $\delta n_{S,d}\gg \delta n_{N,\tau}, \delta n_{N,d}$. Consequently, the contribution to the SDE from Eq.\ \eqref{eq:strong_res_density_corr} is negligible. 

The current $I$ is then given by Eq.\ \eqref{eq:current_main} with $\delta \bar n =0$, and $\delta q_\mathrm{cm}$ determined by Eq.\ \eqref{eq:strong_res_center_of_mass} without the $\delta n_{N,\tau}$ contribution. As a result, the SDE behaves as $\eta \sim \widetilde \eta \sim (\tau\Delta)^{-1}$.

\subsection{Conditions of applicability}
\label{sec:Strong_conditions}

We now examine the conditions of applicability of the obtained solutions. We follow a procedure analogous to that in Sec.\ \ref{sec:Weak_conditions}, focusing on two types of conditions. The first one is when our consideration of a strongly resistive case is applicable, $\theta_{N0} \ll \theta_{S0}$. The second one is when the condition of weak inhomogeneity works, $\delta \theta_{S(N),d} \ll 1$. We should consider the conditions for the $\delta\theta_{S(N),\tau}$ and $\delta \theta_{S(N),d}$ corrections at characteristic frequencies $\omega_n\sim \Delta$ providing the main contribution to the sums in Eqs.\ \eqref{eq:density_zero_order_def}-\eqref{eq:corection_density} for the superfluid densities.

The applicability condition of the strongly-resistive interface case, $\theta_{N0} \ll \theta_{S0}$, can be obtained from Eq.\ \eqref{eq:nontr_corrections}, and amounts to $\tau_N \Delta\gg1$. As discussed in Sec.\ \ref{sec:Weak_conditions}, the quantity $\Delta$ is limited from above by $\Delta_0$ from Eq.\ \eqref{eq:max_Delta_0}. This makes the obtained condition achievable only in the case $\tau_S \gtrsim \tau_N$. The condition is then equivalent to
\begin{equation}
    \tau\Delta\gg 1,
    \label{eq:conditionfull}
\end{equation}
which is written in the same terms as Eq.\ \eqref{eq:appl_weak_11}. (The opposite case, $\tau_S \ll \tau_N$, inevitably implies the limit of weakly nonideal interface, $\tau\Delta \ll 1$.)

The applicability conditions of the weakly inhomogeneous approximation, $\delta \theta_{S(N),d} \ll 1$, follow from Eq.\ \eqref{eq:nontr_corrections}. They have the form 
\begin{equation} \label{eq:condition2}
    d_S\ll \xi_S \sqrt{\tau_S\Delta}, \quad d_N \ll \xi_N,
\end{equation}
which [similarly to Eq.\ \eqref{eq:appl_weak_2}] defines the upper bounds for the thicknesses of the layers. The coherence lengths here are determined by Eq.\ \eqref{eq:coherence_length} with $E_g = 1/\tau_N$. Note that conditions \eqref{eq:condition2} can be fulfilled even for moderately thick layers, $d_{S(N)}\gtrsim \xi_S$, if the interface is sufficiently resistive.

Similarly to Eq.\ \eqref{eq:condition_Thouless}, conditions \eqref{eq:condition2} can be reformulated in terms of the Thouless energies defined in Eq.\ \eqref{eq:Thouless_def}:
\begin{equation} \label{eq:tauEth}
    1/\tau_{S(N)}\ll E_{\mathrm{Th},S(N)}.
\end{equation}
This condition implies that the diffusion times $E_{\mathrm{Th},S(N)} ^{-1}$ are small compared to the escape times $\tau_{S(N)}$. Physically, this means that a quasiparticle diffuses through a layer many times before tunneling through the interface.

Furthermore, the conditions in Eq.\ \eqref{eq:condition2} can be expressed in terms of the interface resistance or transparency: $r_B \gg r_{S(N)}$ or, equivalently, $\mathcal{T} \ll l_{S(N)}/d_{S(N)}$ (here $\mathcal{T}$ is the quantum-mechanical transparency of the interface, and $l_{S(N)}$ is the mean free path in the corresponding layer).

Summarizing, our consideration of the case of weak inhomogeneity and strongly-resistive interface is valid if $\tau_S \gtrsim \tau_N$, for the resistances specified by condition \eqref{eq:conditionfull}, and for the thicknesses specified by Eq.\ \eqref{eq:condition2}.

\section{Limiting cases of the effective AG theory}
\label{sec:AGlimits}

The limiting cases considered in Secs.\ \ref{sec:Weak_case} and \ref{sec:Strong_case}, in the zeroth order reduced to an effective AG theory in each case [see Eqs.\ \eqref{eq:main_equation} and \eqref{eq:main_equation_2}, respectively]. The AG theory admits an analytical description at zero temperature or near the phase transition. In these AG limiting cases, we can obtain the full analytical solution in the sense of finding expressions for $E_g(q_\mathrm{cm},B)$ [or $\Delta(q_\mathrm{cm},B)$] and $\bar n^{(0)}(q_\mathrm{cm},B)$, see Appendix~\ref{Appendix:Abrikosov Gorkov}. If we can also find analytical expressions for the inhomogeneous corrections, we can finally obtain explicit results for $I(q_\mathrm{cm},B)$ and, consequently, for $\eta$ and $\widetilde \eta$.

\subsection{Zero temperature (weakly nonideal interface limit)}
\label{sec:zero_temp}

We start with the case of zero temperature. In this case, we will consider only the limit of small interface resistance, $\tau\Delta \ll1$, which allows full analytical treatment.

Depending on the pair-breaking parameter $\Gamma$, the AG theory [see Eq.\ \eqref{eq:main_equation}] features both gapped and gapless regimes, since the spectral gap behaves as
\begin{equation}
    E_{\mathrm{gap}} =
    \begin{cases}
        E_g \big[1 - \left(\Gamma / E_g\right)^{2/3}\big]^{3/2}, & \Gamma < E_g,\\
        0, & \Gamma > E_g.
    \end{cases}
\end{equation}
In this section, we consider the gapped regime, $\Gamma<E_g(\Gamma)$, since it corresponds to the largest SDE. In this regime, the dependencies of $E_g$ and $\bar n^{(0)}$ on $\Gamma$ are given by Eqs.\ \eqref{eq:order_zero_temperature}-\eqref{eq:g_func} with $z<1$.

A critical current is an extremum of the $I(q_\mathrm{cm})$ function. At the same time, since $E_g(q_\mathrm{cm})$ is a monotonic function, we can maximize/minimize the $I(E_g)$ dependence instead.
Introducing the parameter $x = E_g/E_{g0}<1$, we can write the condition of the gapped regime as $e^{-\pi/4}<x<1$. Expressing the current obtained in Sec.\ \ref{sec:Weak_case} [together with Eq.\ \eqref{eq:current_main}] in terms of $x$, we find 
\begin{gather}
    \frac{I(x,B)}{I_\mathrm{ch}} =  \pm x \left(1 - \frac{16}{3\pi^2} \ln \frac{1}{x}\right) \sqrt{\frac{8}{\pi} x \ln \frac{1}{x} - \frac{B^2}{B_c^2}} + k x^2 \frac{B}{B_c},
    \label{eq:current_zero_temp}
    \\
    \frac{q_\mathrm{cm}(x,B)}{q_c} = \sqrt{\frac{8}{\pi} x \ln \frac{1}{x} - \frac{B^2}{B_c^2}}.
    \label{eq:momentum_zero_temp}
\end{gather}
The upper and lower signs correspond to two opposite directions of the current ($I>0$ and $I<0$, respectively).
The extrema are reached at $x_m$ satisfying the following equation:
\begin{multline}
      \left[\left(\frac{16 + 3\pi^2}{\pi^2}  - \frac{16}{\pi^2}  \ln \frac{1}{x_m} \right)x_m + \frac{4}{3\pi} \frac{B^2}{B_c^2}\right]\ln \frac{1}{x_m} \\
      = x_m + \frac{16+ 3\pi^2}{12\pi} \frac{B^2}{B_c^2} \mp  \frac{\pi k}{2} x_m \frac{B}{B_c} \sqrt{\frac{8}{\pi} x_m \ln \frac{1}{x_m} - \frac{B^2}{B_c^2}} .
    \label{eq:x_eq}
\end{multline}

At zero magnetic field, this equation has the exact solution
\begin{multline}
    x_{m0}=x_m(B=0) \\
    = \exp\left(
    - \frac{16 + 3\pi^2 - \sqrt{256 + 32 \pi^2 + 9 \pi^4}}{32} 
    \right) \approx 0.79,
\end{multline}
leading to
\begin{equation}
    I_{c0} = I(x_{m0},B=0) \approx 0.47 I_\mathrm{ch}.
    \label{eq:zero_T_zero_B_critical}
\end{equation}

According to Eq.\ \eqref{eq:order_zero_temperature}, the $x_{m0}$ value corresponds to the pair-breaking parameter $\Gamma_{m0}/\Gamma_c = -(8/\pi) x_{m0} \ln x_{m0} \approx 0.475$. At same time, the spectral gap disappears at the pair-breaking parameter $\Gamma_{\mathrm{gap}}/\Gamma_c = 2 e^{-\pi/4} \approx 0.91$. This demonstrates that the current in Eq.\ \eqref{eq:zero_T_zero_B_critical} corresponds to the gapped regime. 
Finite magnetic field would then lead to the SDE while still preserving the gapped superconductivity.

The transition to the gapless regime occurs at $x_{\mathrm{gap}} = e^{-\pi/4}$. According to Eq.\ \eqref{eq:x_eq}, this corresponds to the magnetic field
\begin{multline}
    \frac{B_{\mathrm{gap}}}{B_c} = \sqrt{\frac{3 e^{-\pi/4}(16-8\pi+3\pi^2)}{16 - 4\pi +3\pi^2}} \\
    \pm \frac{3 \pi^2 \sqrt{e^{-3\pi/4}(4-\pi)(3\pi-4)}}{(16 - 4\pi + 3 \pi^2 )^{3/2}} k \approx0.92 \pm 0.10 k.
\end{multline}
Since $B_{\mathrm{gap}} \sim B_c$, the range of magnetic fields preserving the gapped superconductivity, $0<B<B_{\mathrm{gap}}$, is quite wide. (At larger magnetic fields, $B>B_{\mathrm{gap}}$, the SDE can be analytically considered in the vicinity of the phase transition, see Sec.\ \ref{Phase_trans} below.)

\begin{figure}[t]
 \includegraphics[width=\columnwidth]{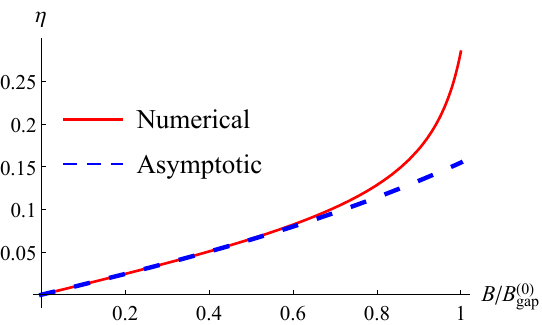}
 \caption{Diode efficiency $\eta$ vs $B$ in the gapped regime at $k=0.1$. The red solid curve results from the numerical solution of Eq.\ \eqref{eq:x_eq}. The blue dashed curve represents the asymptotic solution given by Eq.\ \eqref{eq:approx2}, derived in the limit of small magnetic fields
 ($B \ll B_\mathrm{gap}^{(0)}$). This asymptotic solution provides a good approximation over most of the plotted range. Unlike $\widetilde \eta$, we find that $\eta$ increases monotonically with $B$.}
 \label{fig:contrast_zero}
\end{figure}

\begin{figure}[t]
 \includegraphics[width=\columnwidth]{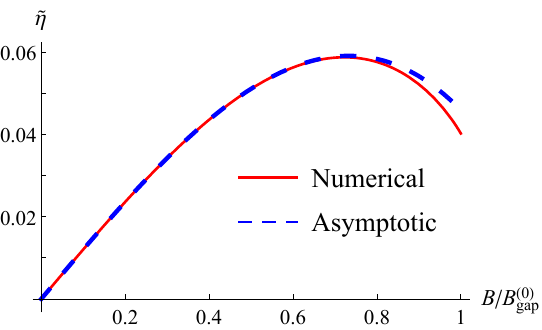}
 \caption{Normalized nonreciprocity $\widetilde \eta$ vs $B$ in the gapped regime at $k=0.1$. The red solid curve results from the numerical solution of Eq.\ \eqref{eq:x_eq}. The blue dashed curve represents the asymptotic solution given by Eq.\ \eqref{eq:approx1}, derived in the limit of small magnetic fields
 ($B \ll B_\mathrm{gap}^{(0)}$). This asymptotic approximation shows excellent agreement with the numerical result across the entire range shown. The maximum of $\widetilde \eta$ is achieved at an intermediate field strength,  $0<B<B_\mathrm{gap}^{(0)}$.}
 \label{fig:absolut_zero}
\end{figure}

To summarize, at zero temperature, the problem reduces to solving Eq.\ \eqref{eq:x_eq}. 
For small magnetic fields, $B\ll B_\mathrm{gap}^{(0)}$, we approximately find the diode efficiency and normalized nonreciprocity [see definitions in  Eqs.\ \eqref{eq:kontrast} and \eqref{eq:absolut_diod}]:
\begin{align}
     \eta(B) &= a k \frac{B}{B_\mathrm{gap}^{(0)}} \Biggl[ 1 + c_1 \biggl( \frac{B}{B_\mathrm{gap}^{(0)}} \biggr)^2 \Biggr],
     \label{eq:approx2}\\
     \widetilde\eta(B) &= a k \frac{B}{B_\mathrm{gap}^{(0)}} \Biggl[ 1 - c_2 \biggl( \frac{B}{B_\mathrm{gap}^{(0)}} \biggr)^2 \Biggr],
     \label{eq:approx1}
\end{align}
where $B_\mathrm{gap}^{(0)}$ is the value of $B_\mathrm{gap}$ at $k=0$ (i.e., in the absence of the SDE).
The numerical constants are expressed as follows:
\begin{align}
    a &= x_{m0}^2 \frac{B_\mathrm{gap}^{(0)}/B_c}{I_{c0}/I_\mathrm{ch}} \approx 1.21,
    \label{eq:a} \\
    c_0 &= \sqrt{256 + 32\pi^2 + 9\pi^4},
    \\
    c_1 &= \frac{(3\pi^2 -16 + c_0) \bigl( B_\mathrm{gap}^{(0)}/B_c \bigr)^2}{6 \pi^{3/2} (I_{c0}/I_\mathrm{ch}) \sqrt{x_{m0}}  \sqrt{16+3 \pi^2 - c_0} } - c_2 \approx 0.28,
    \\
    c_2 &= \frac{\pi(3\pi^2 +16  + c_0) \bigl( B_\mathrm{gap}^{(0)}/B_c \bigr)^2}{12 c_0 x_{m0}} \approx 0.62 .
    \label{eq:c2}
\end{align}
The functions in Eqs.\ \eqref{eq:approx2} and \eqref{eq:approx1} are illustrated in Figs.~\ref{fig:contrast_zero} and~\ref{fig:absolut_zero}, respectively, 
together with numerical results obtained from Eqs.\ \eqref{eq:current_zero_temp} and \eqref{eq:x_eq}.

\begin{figure}[t]
 \includegraphics[width=\columnwidth]{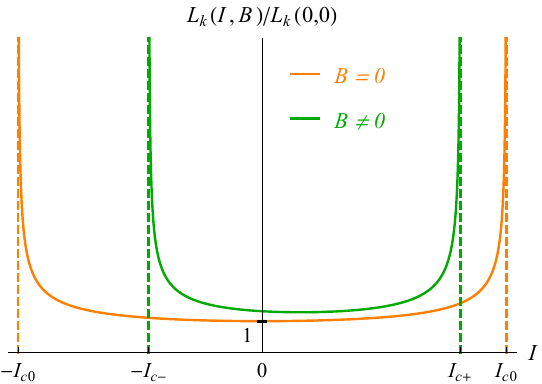}
 \caption{Zero-temperature kinetic inductance $L_k$ vs current $I$ at zero and nonzero magnetic field $B$ [see Eq.\ \eqref{eq:L_zero}] at $k=0.3$. The normalization value is $L_k(0,0) \equiv L_k(I=0,B=0) = L_k(x=1,B=0)$ [note that $I=0$ corresponds to $x=1$ in Eq.\ \eqref{eq:L_zero}]. At $B=0$, the dependence is symmetric, $L_k(I) = L_k(-I)$, and thus the SDE is absent. In contrast, at $B\neq 0$ (the green curve corresponds to $B=0.6 B_c$), this reciprocity is broken, resulting in $L_k(I) \neq L_k(-I)$.}
 \label{fig:L_zt}
\end{figure}

In our case of zero temperature and weakly nonideal interface, we can also discuss the kinetic inductance. In the gapped regime, it can be obtained from Eqs.\ \eqref{eq:current_zero_temp} and \eqref{eq:momentum_zero_temp}:
\begin{equation}
    \frac{L_k(x,B)}{L_k(x=1,B=0)} =  \frac{I_\mathrm{ch}}{q_c}\frac{dq_\mathrm{cm}(x,B)/dx}{dI(x,B)/dx}. 
    \label{eq:L_zero}
\end{equation}
Equation \eqref{eq:L_zero} together with Eq.\ \eqref{eq:current_zero_temp}  parametrically defines the $L_k(I,B)$ dependence (with $x$ being the parameter). This dependence is illustrated in Fig.~\ref{fig:L_zt}, which demonstrates the SDE at $B\neq 0$.

\subsection{Vicinity of the phase transition} 
\label{Phase_trans}

Another case in which explicit answers for the SDE can be obtained, is the vicinity of the superconducting phase transition, i.e., at $\Delta \ll \Delta_0$. This is achieved either near a critical temperature  or near a critical pair-breaking parameter [$T \to T_{c0}$ or  $\Gamma\to \Gamma_c(T)$ in the case of a weakly nonideal interface; $T \to T_{cS}$ or $E_S\to E_{Sc}(T)$ in the case of a strongly resistive interface]. An alternative condition of applicability of this case is discussed in Appendix~\ref{sec:phase_tr_vicinity} [see Eq.\ \eqref{eq:vicinity_condition}].

In both the limiting cases (of weakly and strongly resistive interface), we have previously obtained effective AG equations [see Eqs.\ \eqref{eq:main_equation} and \eqref{eq:main_equation_2}]. In the vicinity of the phase transition, several simplifications take place. 
(i)~The effective order parameter and the superfluid density are determined by Eqs.\ \eqref{eq:gamma_dependence_phase_tr1} and \eqref{eq:gamma_dependence_phase_tr}, respectively (those expressions are immediately applicable in the weakly nonideal case, while in strongly resistive case they work after substitution $E_g\mapsto \Delta$, $\Gamma \mapsto E_S$, and $d_\mathrm{eff} \mapsto d_S$). 
(ii)~In both the cases, the $\delta \bar n$ term in Eq.\ \eqref{eq:current_main} is small and can be neglected. 
(iii)~The effective order parameter and the superfluid density depend on the pair-breaking parameter as $E_g^2\propto \bar n^{(0)} \propto(\Gamma_c-\Gamma)$. Consequently, the $\Gamma$ dependence cancels out from Eq.\ \eqref{eq; delta_q1}, so that $\delta q _{\mathrm{cm}} \propto B$. 

\begin{figure}[t]
 \includegraphics[width=\columnwidth]{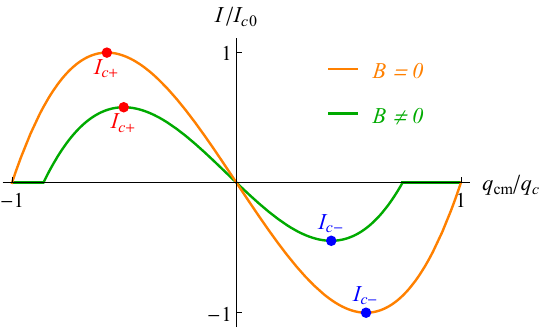}
 \caption{Supercurrent $I$ vs the center-of-mass momentum $q_\mathrm{cm}$ in the absence ($B=0$) and presence ($B \neq 0$) of a magnetic field [see Eq.\ \eqref{eq:current_near_phase_tr}] at $\widetilde k=0.1$. 
 The maxima (red dots) and minima (blue dots) correspond to the critical currents in a positive ($I_{c+}$) and negative ($I_{c-}$) direction, respectively.
 At $B=0$, the dependence is symmetric, $I(-q_\mathrm{cm}) = -I(q_\mathrm{cm})$, and thus the SDE is absent ($I_{c+} = I_{c-}$).  In contrast, at $B\neq 0$ (in the plot, $B=0.6 B_c$), this symmetry is broken, yielding $I(-q_\mathrm{cm}) \neq -I(q_\mathrm{cm})$ and giving rise to the SDE ($I_{c+} \neq I_{c-}$).}
 \label{fig:tok}
\end{figure}

As a result, in both the limiting cases, the expression for the current can be written as
\begin{equation} \label{eq:current_near_phase_tr}
    \frac{I(q_\mathrm{cm},B)}{I_{c0}} = -\frac{3\sqrt{3}}{2} \left[
    1 - \left(\frac{q_\mathrm{cm}}{q_c} + \widetilde k \frac{B}{B_c}\right)^2 - \frac{B^2}{B_c^2}
    \right] \frac{q_\mathrm{cm}}{q_c},
\end{equation}
with different renormalized inhomogeneity parameters:
\begin{equation}
    \widetilde k(T) = \frac{\bar n_0^{(0)}}{\widetilde n(T)} \times
    \begin{cases}
        k \widetilde E_g^2(T)/E_{g0}^2, & \tau\Delta \ll 1,\\
        \bigg[
    \frac{d_N}{d_S} \left(1+\frac{d_N}{d_S}\right) \frac{\sigma_N}{2\sigma_S} \frac{1}{\Delta_0^2 \tau_N^2} \\ \quad \quad + \frac{1}{3 \Delta_0 \tau} \frac{d_S^2}{\xi_S^2} \bigg] \frac{\widetilde\Delta(T)^2}{\Delta_0^2}, & \tau\Delta \gg 1.
    \label{eq:k_near_ph_tr}
    \end{cases}
\end{equation}
Temperature-dependent functions $\widetilde E_g(T)$ and $\widetilde n (T)$ are defined by Eqs.\ \eqref{eq:widetilde_Eg} and \eqref{eq:widetilde_n}, respectively [$\widetilde \Delta(T)$ is defined by Eq.\ \eqref{eq:widetilde_Eg} with substitution $\widetilde E_g \mapsto \widetilde \Delta$ and $\Gamma_c \mapsto E_{Sc}$]. The current is normalized to the critical current at zero magnetic field:
\begin{equation} \label{eq:zero_field_crit_cur}
    I_{c0} = \frac{2}{3\sqrt{3}} \frac{\widetilde n(T)}{\bar n^{(0)}_0} I_\mathrm{ch}.
\end{equation}
The result of Eq.\ \eqref{eq:current_near_phase_tr} is illustrated in Fig.~\ref{fig:tok}.

In the $\tau \Delta \ll1$ case, the inhomogeneity parameter $k$  in Eq.\ \eqref{eq:k_near_ph_tr} is determined by Eq.\ \eqref{eq:full_k1} and contains the $\tau$ and $d$ contributions. In the opposite case, $\tau\Delta \gg1$, the expression in the square brackets in Eq.\ \eqref{eq:k_near_ph_tr} is the inhomogeneity parameter $k$ of the strongly resistive case, also containing $\tau$ and $d$ contributions (the first and the second terms in the square brackets, respectively).
The renormalized inhomogeneity parameter $\widetilde k$ (proportional to $k$) is linearly coupled to $B$ in Eq.\ \eqref{eq:current_near_phase_tr} and thus determines the ``skewed'' $I(q_\mathrm{cm},B)$ dependence and the corresponding strength of the SDE.

Since the current is given by the same Eq.\ \eqref{eq:current_near_phase_tr} in both limiting cases (up to different expressions for $\widetilde k$), we can discuss the critical currents in both cases in a unified manner.
While according to our assumptions, we consider $\widetilde k \ll 1$, the skewed current in the form of Eq.\ \eqref{eq:current_near_phase_tr} is interesting in its own right and allows for some analytical analysis. Below, we perform this analysis exactly whenever possible; otherwise, we retain only the leading nontrivial order with respect to $\widetilde k$.

\begin{figure}[t]
 \includegraphics[width=\columnwidth]{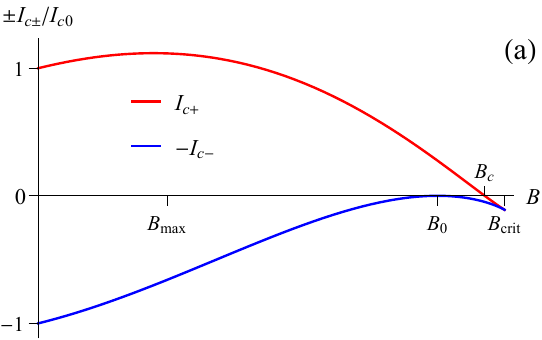}\\[5mm]
 \includegraphics[width=\columnwidth]{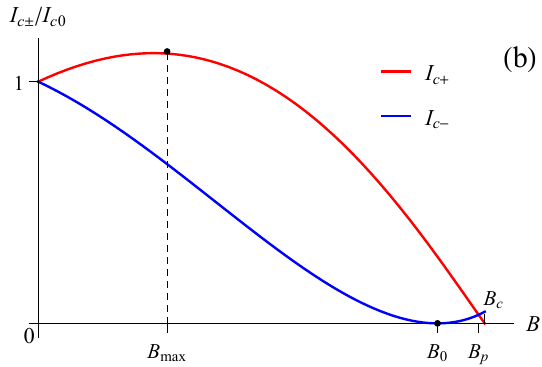}
 \caption{Critical currents $I_{c\pm}$ vs $B$ [see Eq.\ \eqref{eq:crit_currents_near_phase_tr}] at $\widetilde k=0.5$. At $B=0$, the SDE is absent, $I_{c+}(0) = I_{c-}(0)$.
 (a)~Signed critical currents, $\pm I_{c\pm}$.
 As $B$ is increased, $I_{c+}$ reaches a maximum at $B_{\mathrm{max}}$, then decreases and changes its sign at $B_c$. At the same time, $-I_{c-}$ initially behaves monotonically, then reaches zero at $B_0$, and again acquires finite negative values at larger $B$. At $B_{\mathrm{crit}}$ [the actual critical field at finite $\widetilde k$, see Eq.\ \eqref{eq:Bcrit}], the two currents merge, and superconductivity disappears. The intervals $0<B<B_c$ and $B_c<B<B_\mathrm{crit}$ correspond to the SDE and unidirectional superconductivity regimes, respectively.
 (b)~$I_{c\pm}$ in the SDE regime.
 Between $B_0$ and $B_c$, the two curves intersect at $B_p$ [see Eq.\ \eqref{eq:Bp}], where the polarity of the SDE switches. The black dots indicate the maximum and zero points, $B_\mathrm{max}$ and $B_0$. Although the $B_\mathrm{max}$ point is calculated using the leading-order (with respect to $\widetilde k\ll 1$) Eq.\ \eqref{eq:crit_max}, it fits the true position of the maximum well even at not too small value $\widetilde k=0.5$.}
 \label{fig:crittoks_on_B}
\end{figure}

The critical currents are found as extrema of the current as a function of the momentum:
\begin{multline} \label{eq:crit_currents_near_phase_tr}
    \frac{I_{c\pm}(B)}{I_{c0}} =  \frac{1}{3\sqrt{3}} 
    \left(
    \sqrt{3 - (3-\widetilde k^2  ) \frac{B^2}{B_c^2}}
    \pm 2 \widetilde k \frac{B}{B_c}
    \right)
    \\
    \times\left(
    3 - (3+\widetilde k^2) \frac{B^2}{B_c^2}
    \pm \widetilde k \frac{B}{B_c} \sqrt{3 - (3-\widetilde k^2  ) \frac{B^2}{B_c^2}} 
    \right) .
\end{multline}
This result is illustrated in Fig.~\ref{fig:crittoks_on_B}.

At low magnetic fields, the critical currents depend linearly on $B$:
\begin{equation}
    I_{c\pm}(B\ll B_c) \approx  I_{c0} \left(
    1 \pm \sqrt{3} \; \widetilde k B/B_c
    \right).
\end{equation}
As the magnetic field increases, the larger critical current $I_{c+}$ reaches a maximum:
\begin{equation}
    I_{c,\mathrm{max}} \approx I_{c0} \bigl( 1 + \widetilde k^2/2 \bigr)
    \quad \text{at} \quad
     B_{\mathrm{max}} \approx B_c \widetilde k /\sqrt{3}.
    \label{eq:crit_max}
\end{equation}
The smaller critical current $I_{c-}$ also has a special feature; it reaches zero at a magnetic field below $B_c$:
\begin{equation}
     B_0 = 
     B_c \bigl( 1+\widetilde k^2 \bigr)^{-1/2}.
    \label{eq:zanul}
\end{equation}

The behavior at larger $B$ is best explained by Fig.~\ref{fig:crittoks_on_B}(a), which shows the \emph{signed} critical currents, $I_{c+}$ and $-I_{c-}$. At $B>B_0$, the signed critical current $-I_{c-}$ acquires finite negative values again. At $B=B_c$, the signed critical current $I_{c+}$ crosses zero and becomes negative as well. This indicates that we leave the SDE regime and enter the regime of unidirectional superconductivity. In the latter regime, both signed critical currents have the same sign, which implies that superconductivity exists only within an interval of finite currents flowing in the same direction, $-I_{c-}<I<I_{c+}$. In other words, this is a regime of current-induced re-entrant superconductivity (this effect was previously predicted in other systems \cite{Daido2025.PhysRevB.111.L020508,Zhuang2026.SciPostPhys.20.1.021}).
The superconductivity fully disappears at the critical magnetic field
\begin{equation} \label{eq:Bcrit}
B_\mathrm{crit}=B_c  \bigl( 1-\widetilde k^2/3 \bigr)^{-1/2},
\end{equation}
when the two signed critical currents merge. This is the critical field at finite $\widetilde k$, while $B_c$ is the critical value in the absence of the SDE (i.e., at $\widetilde k=0$).

Figure~\ref{fig:crittoks_on_B}(b) demonstrates $I_{c\pm}$ in the interval corresponding to the SDE regime, $0<B<B_c$. The absolute values of the critical currents intersect at
\begin{equation} \label{eq:Bp}
B_p=B_c\bigl( 1+\widetilde k^2/9 \bigr)^{-1/2},
\end{equation}
which implies that the polarity of the SDE changes sign.
(Note that all the characteristic values of $B$ corresponding to the discussed features are determined by the same renormalized inhomogeneity parameter $\widetilde k$ that determines the overall strength of the SDE.)

\begin{figure}[t]
 \includegraphics[width=\columnwidth]{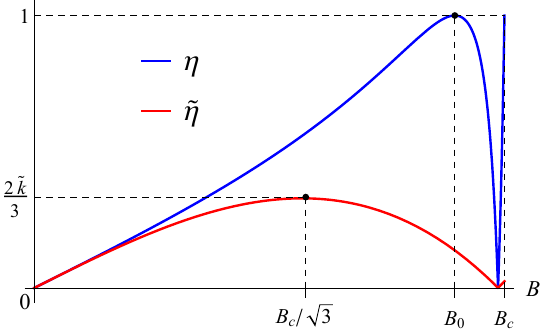}
 \caption{Diode efficiency $\eta$ (blue line) and normalized nonreciprocity $\widetilde \eta$ (red line) vs $B$ in the SDE regime, $0<B<B_c$, [see Eqs.\ \eqref{eq:eta_phase_tr} and \eqref{eq:tilde_eta_phase_tr}] at $\widetilde k =0.5$.
 Both quantities grow linearly at small magnetic fields: $\eta \approx \widetilde{\eta} \propto B$ at $B \ll B_c$, as follows directly from their definitions [see Eqs.\ \eqref{eq:kontrast} and \eqref{eq:absolut_diod}]. Then $\eta$ reaches its maximal value of 1 at $B_0$ (where $I_{c-}=0$) and falls to zero at $B_p$ (where $I_{c+}=I_{c-}$). At the same time, $\widetilde\eta$ reaches its maximum value $\widetilde \eta_{\mathrm{max}}$ at $B_{\widetilde \eta,\mathrm{max}}$, and then likewise decreases to zero at $B_p$. The maximum of $\widetilde\eta$ is well described by the leading terms in Eq.\ \eqref{eq:absolut_max} (black dot at $B_c/\sqrt{3}$) even though we take not too small $\widetilde k$. In the narrow region between $B_p$ and $B_c$, both $\eta$ and $\widetilde\eta$ grow, acquiring finite values again.
 Finally, $\eta$ reaches 1 again at $B_c$, since $I_{c+}$ turns to zero.}
 \label{fig:general}
\end{figure}

The diode efficiency and the normalized nonreciprocity can also be written explicitly:
\begin{align}
    \eta(B) &= \widetilde\eta(B)
    \biggl[
        1- \biggl( 1-\frac{\widetilde k^2}{3} \biggr) \frac{B^2}{B_c^2} 
    \biggr]^{-3/2},
             \label{eq:eta_phase_tr}
             \\
    \widetilde \eta(B) &=
             \sqrt{3}\; \widetilde k \frac{B}{B_c} \biggl|
            1- \biggl(1+\frac{\widetilde k^2}{9}\biggr)\frac{B^2}{B_c^2} 
            \biggr|,
             \label{eq:tilde_eta_phase_tr}
\end{align}
see Fig.~\ref{fig:general} for illustration.

The normalized nonreciprocity has a maximal value at a certain magnetic field:
\begin{equation}
    B_{\widetilde\eta,\max} = \frac{B_c}{\sqrt{3}} \biggl(1+\frac{\widetilde k^2}{9}\biggr)^{-1/2}, \quad
    \widetilde \eta_{\mathrm{max}} = \frac{2 \widetilde k}{3}
    \biggl(1+\frac{\widetilde k^2}{9}\biggr)^{-1/2}.
    \label{eq:absolut_max}
\end{equation}
At the same time, due to the vanishing of one of the critical currents, the diode efficiency is maximal ($\eta = 1$) at $B=B_0$ and $B=B_c$. At $B=B_p$, the polarity of the SDE changes, so at this value $\eta=\widetilde\eta=0$. Since both quantities are positive by definition, they exhibit a V-shaped dependence around $B=B_p$.

\begin{figure}[t]
 \includegraphics[width=\columnwidth]{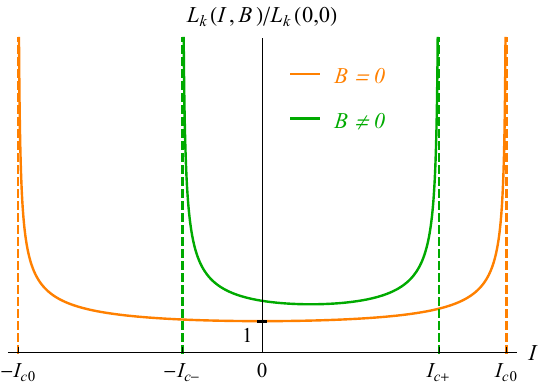}
 \caption{Kinetic inductance $L_k$ vs current $I$ near the phase transition for zero and nonzero magnetic field $B$ [see Eq.\ \eqref{eq:L_near_phase_tr}] at $\widetilde k=0.3$. At $B=0$, the dependence is symmetric, $L_k(I) = L_k(-I)$, and thus the SDE is absent. In contrast, at $B\neq 0$ (the green curve corresponds to $B=0.6 B_c$), the reciprocity is broken, $L_k(I) \neq L_k(-I)$. 
 The orange curve ($B=0$) appears nearly identical to its counterpart in Fig.~\ref{fig:L_zt}, although the underlying functional forms are different. The green curve ($B \neq 0$) is qualitatively similar to the corresponding curve in Fig.~\ref{fig:L_zt} but exhibits a somewhat reduced width at this value of $B$, reflecting smaller critical currents $I_{c\pm}$ relative to the corresponding $I_{c0}$.}
 \label{fig:L_ptr}
\end{figure}

The kinetic inductance in the vicinity of the phase transition is given by relation
\begin{equation}
    \frac{L_k(q_\mathrm{cm},B)}{L_k(0,0)} =  \left(1 - (1+\widetilde k ^2) \frac{B^2}{B_c^2}  - 3 \frac{q_\mathrm{cm}^2}{q_c^2} - 4 \widetilde k \frac{q_\mathrm{cm}}{q_c} \frac{B}{B_c}\right)^{-1}
    \label{eq:L_near_phase_tr}
\end{equation}
together with Eq.\ \eqref{eq:current_near_phase_tr} as a parametric function (with $q_\mathrm{cm}$ being the parameter). The resulting $L_k(I,B)$ dependence is illustrated in Fig.~\ref{fig:L_ptr} which demonstrates the SDE at $B\neq 0$.

\section{Weakly nonideal interface in moderately thick bilayers}
\label{sec:thick}

The regime of a thick bilayer, $d\gtrsim\xi_0$, with a weakly nonideal interface, $\tau\Delta \ll 1 $, is not amenable to full analytical solution, and explicit expressions for the critical currents cannot be derived. Nevertheless, we can obtain estimates in this regime starting from our results for the thin-layer limit, $d\ll\xi_0$, by introducing additional assumptions for thicknesses or conductivities. This would extend the applicability of the results obtained in Sec.\ \ref{sec:Weak_case}.

Below, we assume weak resistance, $\tau\Delta \ll 1$. Our estimates will be based on the results for the inhomogeneity parameters $k_d$ and $k_\tau$ [see Eqs.\ \eqref{eq:k_prozr} and \eqref{eq:full_k}].

\subsection{Strongly contrasting layer thicknesses}

In our previous consideration, we did not make assumptions about the relation between the thicknesses or conductivities of the two layers. Now we consider different cases with respect to the $d_N/d_S$ ratio. 

First, we assume that $d_N/d_S\ll 1$, thereby introducing an additional small parameter. In this limit, the inhomogeneity parameters  can be estimated as
\begin{equation}
    k_d \sim d_S d_N / \xi_0^2, \quad k_\tau \sim (d_N/d_S) \tau\Delta.
\end{equation}
Both the parameters should still be small for the results of Sec.\ \ref{sec:Weak_case} to apply. Now, $k_d$ is small even at $d_S\sim\xi_0$. In this case, the $\tau$ corrections are much smaller: $k_\tau /k_d \sim \tau\Delta \ll1$. The strength of the SDE in this regime is thus determined by the $d$ corrections and does not depend on the interface resistance:
\begin{equation} \label{eq:d_dominance}
    \eta \sim \widetilde \eta  \sim k_d.
\end{equation}

Similarly, we can analyze the opposite case, $d_N/ d_S \gg 1$:
\begin{equation}
    k_d \sim d_S d_N/ \xi_0^2, \quad k_\tau \sim (d_S/d_N) \tau\Delta.
\end{equation}
This limit also implies the dominance of the $d$ corrections, and estimate \eqref{eq:d_dominance} is again valid.
    
Thus, if $d_N/d_S \ll1$ or $d_N/d_S \gg 1$, the interface resistance does not significantly influence the SDE until $\tau\Delta \sim 1$. At larger interface resistances, the induced superconductivity in the N layer becomes suppressed, which weakens the SDE.

Based on these estimates, we anticipate similar behavior for moderately thick bilayers, $d_S, d_N \gtrsim \xi_0$. The contribution of the $\tau$ corrections is then expected to be even more strongly suppressed, with the SDE being predominantly determined by the $d$ corrections. Consequently, the influence of the interface resistance remains negligible until $\tau\Delta\sim 1$, after which it starts to diminish the SDE. In the regime of a strong interface resistance, $\tau\Delta\gg 1$, the $d$ corrections still dominate, so that the SDE decreases as $\eta \sim \widetilde \eta \sim (\tau\Delta)^{-1}$, as discussed in Sec.\ \ref{sec:moderat_thick_l}.

\subsection{Strongly contrasting layer conductivities}

We now examine the case of a moderately thick sample, $d_S \sim d_N \gtrsim \xi_0$, first making an additional assumption about the conductivities of the layers: $\sigma_N /\sigma_S \ll1$. This is the limit of so-called rigid boundary conditions, in which case the derivative $\partial_x\theta_S(0)$ is small [see Eq.\ \eqref{eq:boundary_def1}]. This implies that $\theta_S$ remains nearly constant across the S layer. On the other hand, the superfluid density in the N layer is suppressed, $n_N \propto \sigma_N$, hence $n_N \ll n_S$. The inhomogeneity parameters can then be estimated as
\begin{equation}
    k_{d} \sim \sigma_N/\sigma_S,\quad
        k_{\tau} \sim  \tau\Delta \sigma_N/\sigma_S .
\end{equation}
Similarly, in the opposite case, $\sigma_N/\sigma_S \gg 1$, we obtain
\begin{equation}
    k_{d} \sim \sigma_S/\sigma_N,\quad
    k_{\tau} \sim  \tau\Delta \sigma_S/\sigma_N.
\end{equation}
In both the cases, due to the dominance of the $d$ corrections, Eq.\ \eqref{eq:d_dominance} is again applicable.

\section{Conclusions}
\label{sec:conclusion}

We have analyzed the superconducting diode effect in an SN bilayer in the limiting cases that permit an analytical solution: the case of a weakly nonideal interface with thin layers [see applicability conditions in Eqs.\ \eqref{eq:appl_weak_11} and \eqref{eq:appl_weak_2}] and the case of a strongly resistive interface with thin and moderately thick layers [see applicability conditions in Eqs.\ \eqref{eq:conditionfull} and \eqref{eq:condition2}]. The SDE in this system is of purely orbital nature (due to the orbital influence of the in-plane magnetic field without the Zeeman effect) and is made possible by the spatially inhomogeneous superconducting proximity effect \cite{Levichev2023.PhysRevB.108.094517}. 

We have considered the limiting cases of a weakly nonideal interface, $\tau\Delta \ll 1$ (Sec.\ \ref{sec:Weak_case}), and a strongly resistive interface, $\tau\Delta \gg 1$ (Sec.\ \ref{sec:Strong_case}). In the weakly nonideal case, the bilayer behaves as a single layer with averaged parameters. In the strongly resistive case, superconductivity in the N layer is suppressed, and the bilayer behaves as a single S layer.
In both cases, to leading order, the problem reduces to an effective Abrikosov-Gor'kov theory for homogeneous superconductivity with certain effective parameters [see Eqs.\ \eqref{eq:main_equation} and \eqref{eq:main_equation_2}, respectively]. Pair breaking is caused by the motion of the condensate due to the applied current and magnetic field. The role of the pair-breaking parameter in the AG theory in the two limiting cases is played by the corresponding effective kinetic energy of the condensate, $\Gamma$ [see Eq.\ \eqref{eq:Gamma_def}] or $E_S$ [see Eq.\ \eqref{eq:ES_def}], respectively.
The even (quadratic) dependence of these pair-breaking parameters on the momentum $q_\mathrm{cm}$ and magnetic field $B$ implies that, to leading order, the SDE is absent. We therefore have considered inhomogeneity corrections to the leading-order solution.

\begin{figure}[t]
 \includegraphics[width=\columnwidth]{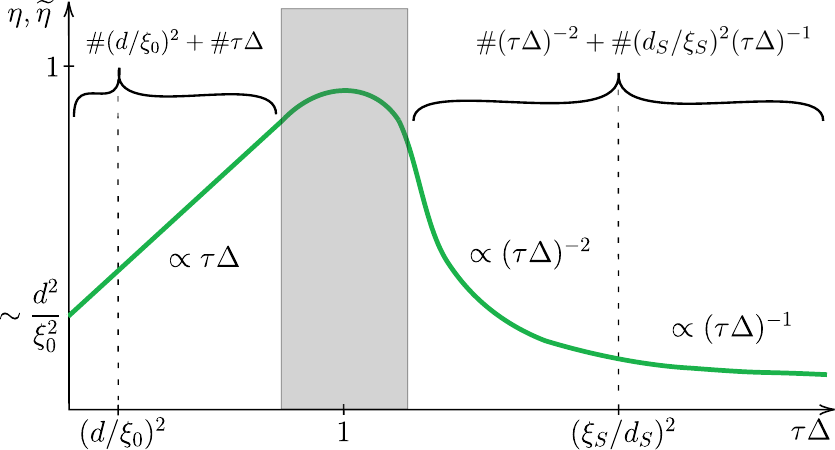}
 \caption{Characteristic value (see text) of the diode efficiency $\eta$ or the normalized nonreciprocity $\widetilde \eta$ vs normalized interface resistance $\tau\Delta$ for a thin bilayer ($d\ll\xi_0$). The shaded region is not amenable to an analytical description.}
 \label{fig:Contrast1}
\end{figure}

\begin{figure}[t]
 \includegraphics[width=\columnwidth]{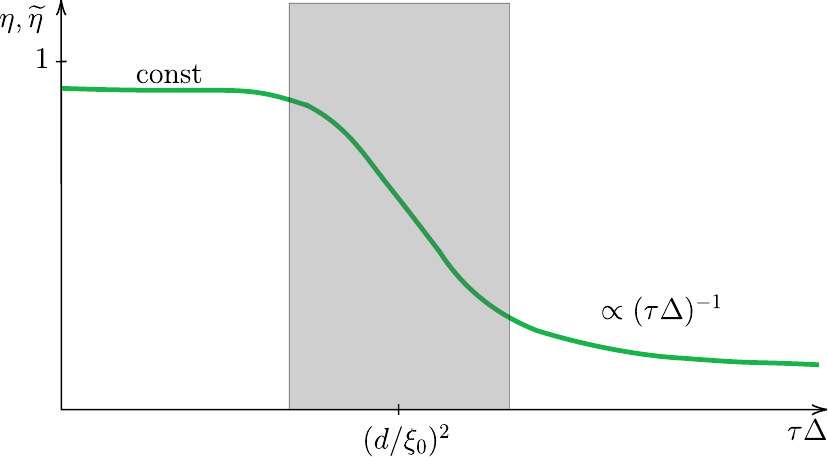}
 \caption{Characteristic value (see text) of the diode efficiency $\eta$ or the normalized nonreciprocity $\widetilde \eta$ vs normalized interface resistance $\tau\Delta$ for a moderately thick bilayer ($d\gtrsim\xi_0$). The shaded region is not amenable to an analytical description.}
 \label{fig:Contrast2}
\end{figure}

We quantify the strength of the SDE using two metrics: the diode efficiency $\eta$ [see Eq.\ \eqref{eq:kontrast}] and the normalized nonreciprocity $\widetilde{\eta}$ [see Eq.\ \eqref{eq:absolut_diod}]. These two quantities behave differently as functions of the magnetic field (see Figs.~\ref{fig:contrast_zero}, \ref{fig:absolut_zero}, and~\ref{fig:general}). 
Both quantities grow linearly at small magnetic fields: $\eta \approx \widetilde{\eta} \propto B$ at $B \ll B_c$. Then $\eta$ reaches its maximal value of 1 at $B_0$ (perfect diode effect) and falls to zero at $B_p$. At the same time, $\widetilde\eta$ reaches its maximum and then likewise decreases to zero at $B_p$. At $B_p$, the polarity of the SDE changes sign. In the narrow region between $B_p$ and $B_c$, both $\eta$ and $\widetilde\eta$ grow again, acquiring finite values (they are positive by definition). Notably, $\eta$ reaches 1 again at $B_c$.
In a narrow interval between the characteristic value $B_c$ and a somewhat larger actual critical field $B_\mathrm{crit}$, a regime of unidirectional superconductivity is realized instead of the SDE.
This means that superconductivity exists only within an interval of finite currents flowing in the same direction (in other words, this is a regime of current-induced re-entrant superconductivity).

At the same time, $\eta$ and $\widetilde \eta$ depend similarly on the interface resistance $\tau\Delta \propto r_B$. Therefore, the dependence of the ``strength'' of the SDE on the interface resistance can be qualitatively characterized by any of these quantities. In Figs.~\ref{fig:Contrast1} and~\ref{fig:Contrast2}, the depicted ``strength'' should be interpreted as the characteristic value of one of these quantities at a field $B\sim B_c$, but not close to the phase transition [for example, one can consider the value of $\eta$ or $\widetilde\eta$ at $B_{\widetilde \eta, \mathrm{max}}$ or $B_c/2$].

In the case of a thin bilayer, $d \ll \xi_0$, two regimes with respect to the interface resistance are realized. As a result, while moderate interface resistance suppresses the proximity effect, it can enhance the SDE (compared to the ideal-interface limit).

In the regime of a weakly nonideal interface, $\tau\Delta\ll 1$ (see Sec.\ \ref{sec:Weak_case}), the Green functions are almost uniform across the whole bilayer. The SDE in this case arises from two types of corrections associated with inhomogeneities: $d$ and $\tau$ corrections, discussed in Secs.\ \ref{sec:d_corr} and \ref{sec:int_corr}, respectively. The strength of the SDE is controlled by the  inhomogeneity parameter $k = k_d + k_\tau$. In this regime, $\eta \sim \widetilde  \eta \sim k$ and increase monotonically. For an ideal interface, the effect is determined by the $d$ correction: $k_d\sim d^2/\xi_0^2$ [see Eq.\ \eqref{eq:k_prozr}]. As the interface resistance increases, a monotonically increasing $\tau$ contribution emerges, scaling as $k_\tau\propto\tau\Delta$ [see Eq.\ \eqref{eq:full_k} and the left part of Fig.~ \ref{fig:Contrast1}]. 

In the regime of a strongly resistive interface, $\tau\Delta\gg1$ (see Sec.\ \ref{sec:Strong_case}), superconductivity is almost suppressed in the N layer while being almost uniform in the S layer. The SDE is then determined by two contributions: the $\tau$ correction in the N layer and the $d$ correction in the S layer [see Eqs.\ \eqref{eq:konzetr_poradki1}, \eqref{eq:konzetr_poradki} and the right part of Fig.~\ref{fig:Contrast1}]. For relatively low resistance, satisfying $\tau\Delta\ll\xi_S^2/d_S^2$, the dominant contribution comes from the superfluid density in the N layer $\delta n_{N,\tau}$, leading to $\eta \sim \widetilde  \eta \sim  \delta n_{N,\tau} \sim (\tau_N \Delta)^{-2}$. In the very high resistance case, $\tau\Delta\gg\xi_S^2/d_S^2$, the SDE arises from the $d$ corrections in the S layer, $\delta n_{S,d}$, leading to $\eta \sim \widetilde \eta \sim \delta n_{S,d} \sim (\tau_S \Delta)^{-1} d_S^2/\xi_S^2$. Consequently, the strength of the SDE decreases monotonically in the strongly resistive regime.

For thin layers, $d \ll \xi_0$, the SDE thus exhibits a nonmonotonic dependence on the interface resistance, with a maximum occurring at a resistance corresponding to $\tau\Delta\sim 1$ (see Fig.~\ref{fig:Contrast1}). This nonmonotonic behavior is prominent in the regime where $\tau$ corrections dominate, specifically when $d^2/\xi_0^2\ll \tau\Delta\ll\xi_S^2/d_S^2$, a regime that exists only in the limit of thin layers. On one hand, a finite-resistance interface decouples the layers and enhances density inhomogeneity, thereby enhancing the SDE. On the other hand, it suppresses superconductivity in the N layer, making the system closer to the limit of the isolated S layer (without the SDE). The competition between these mechanisms gives rise to the nonmonotonic behavior. 

At the same time, in the moderately thick case, $d \gtrsim \xi_0$, the specific regime of nonmonotonicity disappears. The corresponding Fig.~\ref{fig:Contrast2} can be obtained from Fig.~\ref{fig:Contrast1} by ``compressing'' the area between the two vertical dashed lines (as $d$ becomes of the order of $\xi_0$ and $\xi_S$). In the regime of large interface resistances (now defined by the condition $\tau\Delta \gg d^2/\xi_0^2$), the system's behavior is indistinguishable from the corresponding regime of the thin-bilayer limit, leading again to $\eta \sim \widetilde \eta \sim (\tau_S \Delta)^{-1}  d_S^2/\xi_S^2$. At the same time, in the low interface resistance regime, $\tau\Delta \ll d^2/\xi_0^2$, the SDE is approximately constant as a function of the interface resistance,
$\eta, \widetilde \eta=\mathrm{const}_\tau$.

The characteristic behavior of $\eta$ and $\tilde\eta$ as functions of the normalized thickness $d/\xi_0$ can be inferred from the results presented in
Figs.~\ref{fig:Contrast1} and~\ref{fig:Contrast2}. For definiteness, we consider varying the total thickness $d$ while keeping the ratio of the layer thicknesses fixed.
In the limit of a weakly nonideal interface, $\tau\Delta\ll 1$, both $\eta$ and $\tilde\eta$ increase from small values of order $\tau\Delta$ to values of order unity at $d/\xi_0\sim 1$, and remain at this level as the thickness is increased further.
In the limit of a strongly resistive interface, $\tau\Delta\gg 1$, a similar increase occurs, but from values of order $(\tau\Delta)^{-2}$ to values of order $(\tau\Delta)^{-1}$, which nevertheless remain small in absolute magnitude.

We also characterize the SDE in terms of the kinetic inductance $L_k$ in the limiting cases of zero temperature (Sec.\ \ref{sec:zero_temp}) and the vicinity of the phase transition (Sec.\ \ref{Phase_trans}). The SDE manifests itself as a nonreciprocal current-dependent kinetic inductance, $L_k(I) \neq L_k(-I)$ (see Figs.~\ref{fig:L_zt} and~\ref{fig:L_ptr}). The dependencies $L_k(I,B)$ are qualitatively similar in both cases. 

Experimentally, a pronounced SDE with relatively large currents can be observed in moderately thick bilayers with ideal interfaces. At the same time, in thin bilayers the largest SDE is realized at finite values of the interface resistance. Since our consideration implies $l \ll d \ll \lambda$, the range of the  thicknesses $d$ is from nanometers to microns.

\begin{acknowledgments}
We thank D.~Yu.\ Vodolazov for useful discussions.
The work was supported by the Russian Science Foundation (Grant No.\ 24-12-00357).
\end{acknowledgments}

\appendix

\section{Kinetic inductance}
\label{Appendix:Kinetic inductance}

In this Appendix, we derive Eq.\ \eqref{eq:kinetic_appearance} for the kinetic inductance $L_k$ and discuss its relevance to experiment. Generally, we consider an arbitrary thin strip of thickness $d \ll \lambda$ and width $w$, confined to the region $0<x<d$, whose properties may be inhomogeneous across the thickness
(e.g., a layered structure with superconducting and normal layers). Let $n_x(q)$ denote the superfluid density at coordinate $x$ with momentum $q$. Kinetic inductance in inhomogeneous systems with fixed spatial distribution $n_x$ (reflecting intrinsic properties of layered materials) has been studied in a number of works, see, e.g., Refs.\ \cite{Claassen1991.PhysRevB.44.9605, Neilo2025}.

However, following Ref.\ \cite{Levichev2023.PhysRevB.108.094517}, we want to take into account the nonlinearity implying that $n_x(q)$ depends on momentum (i.e., on the currents flowing in the structure). In the presence of a magnetic field, $n_x$ at each coordinate depends on its own momentum $q_s(x)$. In the thin-layer case, this momentum profile is given by Eq.\ \eqref{eq:super_velosity_definition} with some $q_0$ to be determined from the condition of fixed total current:
\begin{equation}
    I(q_0) = -\frac{ew}{2m} \int_0^{d} n_x\big(q_s(x)\big) q_s(x) dx.
    \label{eq:current_on_momentum}
\end{equation}

\subsection{Derivation}

The kinetic inductance is defined through the relation between the kinetic energy per
unit length and the total current. In the simplest linear case and in the absence of a magnetic field, this relation takes the form:
\begin{equation}
    E_k = L_k I^2 / 2.
    \label{eq:kin_energy_linear}
\end{equation}
Linearity here implies that the density does not depend on the current or momentum, i.e., $n_x = \mathrm{const}_q$, leading to a current-independent inductance $L_k = \mathrm{const}_I$. In a general nonlinear case, Eq.\ \eqref{eq:kin_energy_linear} should be generalized as
\begin{equation}
    E_k(I) = \int_0^I L_k(I) I dI.
    \label{eq:kin_energy_nonlinear}
\end{equation}

To obtain the kinetic inductance from Eq.\ \eqref{eq:kin_energy_nonlinear}, we express the kinetic energy in terms of the momenta of Cooper pairs with the mass $2m$ and the density $n_x/2$. For a given coordinate $x$, a change in the local momentum by $dq$ contributes an energy increment $n_x(q) q dq /4m$. Here, we explicitly allow the number of superconducting electrons to depend on the momentum itself. Integrating over momenta and coordinates, we obtain the total kinetic energy of the entire layer:
\begin{equation}
    E_k^{\mathrm{tot}}(q_0) = \frac{w }{4m}\int_0^d dx \int_{0}^{q_{0}-2 e B x} n_x(q) q dq.
    \label{eq:tot_kin_energy_momentum}
\end{equation}
This energy contains two contributions: from the transport current $I$ and from ``vortex'' currents generated by the magnetic field $B$.

First, if we apply the magnetic field $B$ even in the absence of the transport current $I$, ``vortex'' currents are generated, which are inhomogeneous across the thickness of the strip. 
They correspond to the magnetic part of the total kinetic energy:
\begin{equation}
    E_k^B = E_k^{\mathrm{tot}}(q_{00}).
    \label{eq:B_kin_energy_momentum}
\end{equation}
Here, the momentum $q_{00}$ corresponds to the state with zero transport current, $I(q_{00})=0$. 

If we now turn on the transport current $I$, the magnetic part of the kinetic energy, Eq.\ \eqref{eq:B_kin_energy_momentum}, does not change. At the same time, we obtain an additional energy contribution
\begin{equation}
    E_k(q_0) = E_k^{\mathrm{tot}}(q_0) - E_k^B  = \frac{w }{4m}\int_0^d dx \int_{q_{00}-2 e B x}^{q_{0}-2 e B x} n_x(q) q dq.
    \label{eq:kin_energy_momentum}
\end{equation}
This is the transport-current part of the total kinetic energy, which is related to the kinetic inductance and is equivalent to Eq.\ \eqref{eq:kin_energy_nonlinear}. Equations \eqref{eq:kin_energy_momentum} and \eqref{eq:current_on_momentum} thus parametrically define the $E_k(I)$ function in Eq.\ \eqref{eq:kin_energy_nonlinear}.

Differentiating Eq.\ \eqref{eq:kin_energy_momentum} with respect to $q_0$, we obtain the same integral as in Eq.\ \eqref{eq:current_on_momentum}:
\begin{equation}
    \frac{dE_k(q_0)}{dq_0} = \frac{w}{4m} \int_0^d  n_x\big(q_s(x)\big) q_s(x) dx = - \frac{1}{2e} I(q_0).
\end{equation}
Integrating this relation backwards, we can write
\begin{equation}
    E_k(q_0) = -\frac{1}{2e} \int_{q_{00}}^{q_0} I(q_0) d q_0 = - \frac{1}{2e} \int_0^I I \frac{dq_0}{dI} dI.
\end{equation}
Comparing this to Eq.\ \eqref{eq:kin_energy_nonlinear}, we finally obtain Eq.\ \eqref{eq:kinetic_appearance}.

Although $q_0 = q_s(0)$ is defined as the momentum at a specific reference point $x=0$, the final result in Eq.\ \eqref{eq:kinetic_appearance} is independent of this choice. We can actually substitute $q_0$ in Eq.\ \eqref{eq:kinetic_appearance} by an arbitrary $Q= q_s(x')$ at any specific fixed point $x'$ without altering the expression [note that $dQ = dq_0$ according to Eq.\ \eqref{eq:super_velosity_definition}]. At the same time, although physically transparent, the center-of-mass momentum $q_\mathrm{cm}$ is inappropriate for this substitution because the center of mass of the superconducting condensate is not a fixed point, $x_\mathrm{cm} \neq \mathrm{const}_{I,B}$.

\subsection{Relation to experiment}

We can rewrite Eq.\ \eqref{eq:kin_energy_nonlinear} for the kinetic energy as a work for acceleration of superconducting electrons with power $I \mathcal{E}$ during time $t$:
\begin{equation}
    E_k = \int_0^t I \mathcal{E} dt,\qquad
    \mathcal{E} = L_k \frac{dI}{dt}.
\end{equation}
Here, $\mathcal{E}$ is the electric field (voltage per unit length) that accelerates electrons. The relation between $\mathcal{E}$ and $I$ corresponds to the kinetic contribution to the impedance $Z_k =  -i\omega L_k$ at frequency $\omega$. The kinetic contribution together with the contribution $Z_g = -i \omega L_g$ from the geometric inductance $L_g$ compose the total impedance $Z = Z_k + Z_g$ that can be measured in experiment \cite{Levichev2023.PhysRevB.108.094517}.

\section{Abrikosov-Gor'kov (AG) theory}
\label{Appendix:Abrikosov Gorkov}

The limiting cases considered in Secs.\ \ref{sec:Weak_case} and \ref{sec:Strong_case}, in the zeroth order reduced to an effective AG theory in each case [see Eqs.\ \eqref{eq:main_equation} and \eqref{eq:main_equation_2}, respectively] with the effective order parameter $E_g$ and $\Delta$, and with the pair-breaking parameter $\Gamma$ and $E_S$, respectively. The AG theory admits an analytical description at zero temperature or in the vicinity of the phase transition \cite{Abrikosov1960,Maki1969}. In this Appendix, we present analytical results for the effective order parameter and average superfluid density as functions of the pair-breaking parameter in these two limits. 

For definiteness, our notations will correspond to the $\tau\to 0$ limit of Sec.\ \ref{sec:Weak_case}. The results in the $\tau\to \infty$ limit of Sec.\ \ref{sec:Strong_case} can be obtained by a formal substitution $\Gamma \mapsto E_S$, $E_g\mapsto \Delta$, $d_\mathrm{eff} \mapsto d_S$, and $T_{c0}\mapsto T_{cS}$.

\subsection{Effective self-consistency equation}

Consideration in Sec.\ \ref{sec:Weak_case}, which led to Eq.\ \eqref{eq:main_equation}, should be supplemented by discussion of the self-consistency equation. Its effective form can be obtained from Eq.\ \eqref{eq:self_consist_initial} with homogeneous $\theta_S(x) = \theta$ and $\Delta(x) = \Delta$. In terms of the effective order parameter $E_g$ [see Eq.\ \eqref{eq; def_minigap}], the effective self-consistency equation can be written as
\begin{equation}
    E_g \ln \frac{T}{T_{c0}} = 2 \pi T \sum_{\omega_n>0} \left(\sin \theta(\omega_n) - \frac{E_g}{\omega_n}\right).
    \label{eq:self_consist_new}
\end{equation}
Here,  $T_{c0}$ is the critical temperature of the bilayer as a whole in the absence of current and magnetic field ($\Gamma =0$). It is related to the effective order parameter by the standard BCS relation
\begin{equation}
    T_{c0} = ( e^\gamma / \pi ) E_{g0},
    \label{eq:Tc0}
\end{equation}
where $E_{g0} = E_g(\Gamma=0,T=0)$ is the zero-temperature order parameter in the absence of a current and magnetic field.

Thus, the problem is reduced to solving Eq.\ \eqref{eq:main_equation} together with Eq.\ \eqref{eq:self_consist_new}. This can be done analytically at zero temperature (see Appendix~\ref{sec:Apendix_zero_temp}) or near the phase transition, $E_g \to 0$ (see Appendix~\ref{sec:phase_tr_vicinity}).

\subsection{Critical pair-breaking parameter}
\label{sec:critical_Gamma}

Equation \eqref{eq:main_equation} is valid for the pair-breaking parameter less than the critical value $\Gamma_c(T)$. At the critical value, a phase transition occurs and superconductivity disappears: $\theta(\Gamma\to\Gamma_c) \to 0$ and $E_g(\Gamma\to\Gamma_c) \to 0$. In this limit, Eq.\ \eqref{eq:self_consist_new} produces the following equation for the critical pair-breaking parameter:
\begin{equation}
    \ln \frac{T_{c0}}{T} = \psi\left(\frac{1}{2} + \frac{\Gamma_c(T)}{2\pi T}\right) - \psi\left(\frac{1}{2}\right).
    \label{eq:critical_pairbreaking}
\end{equation}
Here, $\psi$ denotes the digamma function. 

In the limits of zero temperature ($T=0$) and near the critical temperature ($T_{c0} -T \ll T_{c0}$), we obtain
\begin{equation}
    \Gamma_c(T) = \begin{cases}
        \displaystyle E_{g0}/2, & T=0,\\
        \displaystyle 4(T_{c0}-T)/\pi, & T\to T_{c0}.
    \end{cases}
    \label{eq:critical_pairbreaking_asimp}
\end{equation}

\subsection{Zero-temperature limit}
\label{sec:Apendix_zero_temp}

At zero temperature, we can replace the sum in Eq.\ \eqref{eq:self_consist_new} with an integral and calculate it explicitly. We obtain the dependence of the effective order parameter on the pair-breaking parameter, $E_g(\Gamma)$, determined by the following equation:
\begin{equation}
    \ln (E_g/E_{g0}) + f(\Gamma/E_g)=0,
    \label{eq:order_zero_temperature}
\end{equation}
with the $f$ function defined as
\begin{equation}
    f(z) = \begin{cases}
        \displaystyle \pi z/4, & z<1,\\
        \displaystyle \arccosh z + \frac{z}{2} \arcsin \frac{1}{z} - \frac{\sqrt{z^2-1}}{2z}, & z>1.
    \end{cases}
\end{equation}
The point $E_g(\Gamma) = \Gamma = e^{-\pi/4} E_{g0}$ (i.e., $z=1$), where the behavior of $E_g(\Gamma)$ changes qualitatively, corresponds to the transition to the gapless regime.

We also obtain the average superfluid density $\bar n^{(0)}$ by similarly reducing the sum in Eq.\ \eqref{eq:density_zero_order_def} to an integral. As a result, we get
\begin{equation}
    \frac{\bar n^{(0)}(\Gamma)}{\bar n_0^{(0)}} = \frac{E_g(\Gamma)}{E_{g0}} g\left(\frac{\Gamma}{E_g(\Gamma)}\right),
    \label{eq:density_zero_temperature}
\end{equation}
with the $g$ function defined as
\begin{multline}
    g(z) = \\ \begin{cases}
        \displaystyle 1 - 4z/3\pi, & z<1,\\
        \displaystyle\frac{2}{\pi}\left\{
        \arcsin \frac{1}{z} - \frac{z}{3} \left[ 2 - \left(2+ \frac{1}{z^2}\right)\sqrt{1 - \frac{1}{z^2}} 
        \right]
        \right\}, & z>1.
    \end{cases}
    \label{eq:g_func}
\end{multline}

\subsection{Vicinity of the phase transition}
\label{sec:phase_tr_vicinity}

In the vicinity of the phase transition ($\theta\to0$, $E_g\to 0$), we can find an explicit function $\theta(\omega_n>0)$ from Eq.\ \eqref{eq:main_equation} and solve the self-consistency Eq.\ \eqref{eq:self_consist_new}:
\begin{gather}
    \theta(\omega_n) \approx\frac{E_g}{\Gamma+\omega_n} - \frac{E_g^3 \omega_n}{2(\Gamma +\omega_n)^4},
    \\
    E_g^2(T,\Gamma) = \widetilde E_g^2(T) \bigl( 1 - \Gamma/\Gamma_c(T) \bigr).
    \label{eq:gamma_dependence_phase_tr1}
\end{gather}
Here, the temperature-dependent prefactor is given by
\begin{equation}
    \widetilde E_g^2(T) =T \Gamma_c(T) \Xi \left(\frac{\Gamma_c(T)}{2 \pi T}\right),
    \label{eq:widetilde_Eg}
\end{equation}
where $\Xi$ denotes the following positively defined function [$\Xi(x)>0$]:
\begin{equation}
    \Xi(x) = -\frac{ 24 \pi \psi'(1/2+x)}{3 \psi''(1/2+x) + x \psi'''(1/2+x)}.
    \label{eq:big_psi}
\end{equation}
In the limiting cases of zero temperature ($T=0$) and near the critical temperature ($T_{c0} -T \ll T_{c0}$), the effective order parameter behaves as follows:
\begin{equation}
    E_g^2 = \begin{cases}
        \displaystyle6 E_{g0} \bigl( \Gamma_c(0) - \Gamma\bigr), & T =0,\\
        \displaystyle\frac{2 \pi^3 T_{c0}}{7 \zeta(3)} \bigl( \Gamma_c(T) - \Gamma \bigr), & T\to T_{c0}.
    \end{cases}
\end{equation}
The critical pair-breaking parameter $\Gamma_c(T)$ entering these equations, in the same limiting cases was found in Eq.\ \eqref{eq:critical_pairbreaking_asimp}. 

Equation \eqref{eq:gamma_dependence_phase_tr1} allows us to rewrite condition $E_{g0} \ll E_g$ (vicinity of the phase transition) in terms of proximity of the pair-breaking  parameter to the critical value:
\begin{equation}
    \Gamma_c(T) - \Gamma \ll E_{g0}.
    \label{eq:vicinity_condition}
\end{equation}
This condition can be satisfied not only in the obvious limit of $\Gamma \to \Gamma_c$, but also in the limit of $T\to T_{c0}$, in which case $\Gamma_c(T) \ll E_{g0}$ and $\Gamma$ can take arbitrary values between $0$ and $\Gamma_c(T)$.

Finally, we use Eq.\ \eqref{eq:gamma_dependence_phase_tr1} to obtain the superfluid density:
\begin{equation}
    \bar n^{(0)}(T,\Gamma) = \widetilde  n(T) \bigl( 1 - \Gamma/\Gamma_c(T) \bigr),
    \label{eq:gamma_dependence_phase_tr}
\end{equation}
with the temperature-dependent prefactor
\begin{equation}
    \frac{\widetilde  n(T)}{\bar n_0^{(0)}} = \frac{\widetilde E_g^2(T)}{2 \pi^2 T E_{g0}} \psi'\left(\frac{1}{2} + \frac{\Gamma_c(T)}{2\pi T}\right).
    \label{eq:widetilde_n}
\end{equation}



%

%
%
%
%
%

\end{document}